\documentclass[aps,preprint,onecolumn,groupedaddress,nofootinbib,superscriptaddress]{revtex4-2}
\usepackage{graphicx}  
\usepackage{dcolumn}   
\usepackage{bm}        
\usepackage{amsmath,amssymb,amsxtra,bm,epsfig}   
\usepackage{float}
\usepackage[dvipsnames]{xcolor}
\usepackage{xspace}
\usepackage{pifont}
\usepackage{soul}
\usepackage[]{changes}

\usepackage[bookmarks, breaklinks, colorlinks,urlcolor=RoyalBlue, citecolor=RubineRed, 
linkcolor=ForestGreen]{hyperref}

\usepackage[font=scriptsize]{subcaption}

\begin{document}
	
	
\title{Phenomenology of the flavor symmetric scoto-seesaw model with  dark matter and TM$_1$ mixing}
\author{Joy Ganguly}
\email{joyganguly.hep2022@gmail.com}
\affiliation{Department of BSH, University of Engineering and Management, Kolkata, India}

\author{Janusz Gluza}
\email{janusz.gluza@us.edu.pl}
\affiliation{Institute of Physics, University of Silesia,  Katowice, Poland}

\author{Biswajit Karmakar}
\email{biswajit.karmakar@us.edu.pl}
\affiliation{Institute of Physics, University of Silesia,  Katowice, Poland}

\author{Satyabrata Mahapatra}
\email{satyabrata@g.skku.edu}
\affiliation{Department of Physics and Institute of Basic Science, Sungkyunkwan University, Suwon 16419, Korea}
	
	
	\begin{abstract}
	  We propose a hybrid scoto-seesaw model based on the $A_4 \times Z_4 \times Z_3 \times Z_2$ non-Abelian discrete flavor symmetry. Light neutrino masses come from the tree-level type-I seesaw mechanism  and from the one-loop scotogenic contribution accommodating viable dark matter candidates responsible for observed relic abundance of dark matter (DM). Respectively, both these contributions restore the atmospheric and solar neutrino mass scales. With only one right-handed neutrino, the model features specific predictions with the normal ordering of light neutrino masses, the lightest neutrino being massless, only one relevant CP Majorana phase. Further, experimentally favorable TM$_1$ mixing scheme is realized with concrete correlations and constraints on the mixing angles and associated CP phases. The model predicts the atmospheric mixing angle to be in the upper octant with specific ranges  $0.531 (0.580) \leq \sin^2\theta_{23}\leq 0.544 (0.595) $,  and the Dirac CP phase is restricted within the range $\pm(1.44-1.12)$ radian. The Majorana phase is also tightly constrained with a range of $0.82-0.95$ and $1.58-1.67$ radian, which is otherwise unconstrained from neutrino oscillations.  Strict predictions on the Majorana phases also yield an accurate prediction for the effective mass parameter for neutrinoless double beta within the range of $1.61-3.85$ meV.  The model offers a rich phenomenology regarding DM relic density and direct search constraints,  and the fermionic DM scenario has been discussed in detail, estimating its possible connection with the neutrino sector.  As an example of the model studies at colliders, the SM Higgs in the diphoton decay channel is examined. The model predicts strictly vanishing $\tau \to e\gamma$, $\tau \rightarrow 3e$ decays and testable signals by  MEG-II and SINDRUM/Mu3e experiments for the $\mu \to e \gamma$  and $\mu \to 3 e$ decays, respectively.
	\end{abstract}

	\maketitle

\section{Introduction}\label{sec:intor}
In the last few decades, several experiments around the globe have confirmed the phenomenon of neutrino oscillation with incredible precision~\cite{Fukuda:2001nk, Ashie:2005ik, Ahmad:2002jz, Ahmad:2002ka, Abe:2008aa, Abe:2011sj, Abe:2011fz, Abe:2013hdq}. The immediate consequence of neutrino oscillation is that at least two light neutrinos have nonzero mass. Furthermore, if we combine this with a bound on the absolute neutrino masses coming from the end-point spectrum of the tritium beta decay~\cite{KATRIN:2021uub}, as well as the bounds from cosmological surveys~\cite{Planck:2018vyg} and the neutrinoless double beta decay~\cite{KamLAND-Zen:2022tow}, we conclude that the neutrino masses are in the sub-eV scale. Despite these spectacular accomplishments, the origin of tiny neutrino masses (compared to other Standard Model fermions)  remains an open question in particle physics. Over the years various ideas have been proposed, and the most common schemes are seesaw mechanisms~~\cite{Minkowski:1977sc, GellMann:1980vs, Mohapatra:1979ia, Mohapatra:1986aw, Magg:1980ut, Lazarides:1980nt, Schechter:1980gr, Mohapatra:1986bd} and radiative generation of neutrino masses~\cite{Cai:2017jrq}. Nonzero neutrino masses can also be realized within the framework of hybrid mass mechanisms where both seesaw and radiative mass mechanisms contribute.

In addition to the tiny neutrino masses, we are yet to understand the observed pattern of the lepton mixing comprehensively. In fact, two of the three mixing angles, namely solar ($\theta_{12}$) and atmospheric ($\theta_{23}$), are found to be large, while the reactor ($\theta_{13}$)  mixing angle is relatively small. Such a finding clearly shows the distinctive feature associated with the lepton sector in contrast to the quark sector.  The study of the underlying principle behind this typical mixing is particularly interesting with the precise measurement of the reactor mixing angle $\theta_{13}$ \cite{Abe:2011fz, An:2012eh, Ahn:2012nd, Abe:2013hdq}. Neutrino oscillation data also constraints the two mass squared differences (solar and atmospheric) defined as $\Delta m_{21}^2 = m_2^2-m_1^2$ and $|\Delta m_{31}^2| = | m_3^2-m_1^2|$ where $m_1, m_2, m_3$ are the masses of the three light neutrinos.  The present global analysis from
several experimental data can be summarized as~ \cite{Esteban:2020cvm}
\begin{eqnarray}
&&  \Delta m^2_{21}=(6.82 - 8.03)\times10^{-5}\hspace{.1cm} \rm{eV}^2, \hspace{.5cm} |\Delta m^2_{31}|=(2.427 - 2.590)\times10^{-3}\hspace{.1cm} \rm{eV}^2, \nonumber  \\
 && \sin^2\theta_{12}=0.270-0.341, \hspace{.2cm} \sin^2\theta_{23}=0.408-0.603, \hspace{.2cm} \sin^2\theta_{13}=0.02052-0.02398,  \label{eq:oscdata}
\end{eqnarray}
for normal ordering (NO) of light neutrino mass and similar constraints for inverted ordering (IO)~\cite{Esteban:2020cvm}.
In this regard, many conjectures have been put forward. A particular pattern yielding $\sin^2 \theta_{23} = 1/2, \sin^2  \theta_{12} = 1/3$ and $\theta_{13}=0$ known as tri-bimaximal mixing  (TBM)~\cite{Harrison:2002er},  received a lot of attention due the proximity of $\theta_{23}$ and $\theta_{12}$ with experimental values. Such a  mixing pattern can also be elegantly generated using flavor symmetries. Particularly, the use of non-Abelian discrete symmetries like $S_3, A_4, S_4, A_5$ is very well known~\cite{Ma:2001dn,Altarelli:2005yx} in this context. For a detailed discussion on such frameworks see~\cite{Ishimori:2010au, Altarelli:2010gt, King:2013eh, Petcov:2017ggy, Chauhan:2022gkz,Chauhan:2023faf} and references therein. Unsurprisingly, a  deformation from TBM mixing becomes inevitable after precisely measuring $\theta_{13}$. Nevertheless, the TBM mixing scheme can still be considered as a leading-order approximation, requiring adjustments such as accounting for non-zero $\theta_{13}$ and  the Dirac CP phase $\delta$. Possible simple deviations from the TBM mixing are called  trimaximal (TM$_1$ and TM$_2$) mixings where the first and second columns of the TBM mixing respectively remain identical~\cite{Albright:2008rp,Bjorken:2005rm, Albright:2010ap, He:2006qd, Luhn:2013lkn, deMedeirosVarzielas:2012apl, Shimizu:2017fgu, Thapa:2021ehj, Chakraborty:2020gqc}. Such deviations can be elegantly achieved by considering larger residual symmetry (compared to the TBM scenario)   or introducing an additional constituent that breaks the TBM structure~\cite{Thapa:2021ehj, Chakraborty:2020gqc, deAdelhartToorop:2011nfg, deAdelhartToorop:2011re, Hernandez:2012sk,Ding:2012xx, Karmakar:2014dva, Karmakar:2015jza, Karmakar:2016cvb, Karmakar:2017sxt}. Although the $3\sigma$ allowed ranges for all three mixing angles can be explained by both TM$_1$ and TM$_2$ mixings, the allowed value of the  solar mixing angle $\theta_{12}$ (within these trimaximal scenarios)  slightly prefers the TM$_1$ over the TM$_2$ mixing scheme.  For a detailed discussion on the relative comparison of both mixings, see ~\cite{Chauhan:2023faf}.

Apart from the neutrino masses and mixing, unraveling the nature of dark matter (DM) remains a pressing challenge in contemporary particle physics. While compelling astrophysical evidence, including observations like galaxy rotation curves, gravitational lensing, and the cosmic large-scale structure, substantiates the existence of DM~\cite{Cheung:2013dua}, the quest for a laboratory-based confirmation persists. Satellite missions such as WMAP~\cite{WMAP:2012nax} and PLANCK~\cite{Aghanim:2018eyx} have precisely determined that DM constitutes roughly $26.8\%$ of the total energy content of the Universe.  Expressing the prevailing dark matter abundance through the density parameter $\Omega_{\rm DM}$ and the normalized Hubble parameter $h$ (Hubble Parameter divided by $100$ km $s^{-1}Mpc^{-1}$) yields $\Omega_{\rm DM}h^{2} = 0.120\pm0.001$ at a $68\%$ confidence level. Still the intricacies of its properties beyond gravitational interactions remain elusive. Among all proposed particle dark matter, the most sought-after paradigm is the weakly interacting massive particle (WIMP) paradigm, which suggests a dark matter particle with a mass and interaction strength akin to the electroweak scale.
Unfortunately, the Standard Model of particle physics fails to comprehensively explain neutrino masses, mixings and dark matter. Standing at this juncture, certainly, it is a tempting challenge to find a common origin of these two seemingly uncorrelated sectors, if any. Hence, we aim to go beyond the SM of particle physics to explore scenarios that can accommodate a candidate of DM and explain non-zero neutrino masses and mixings.
 
Neutrino oscillation data presented earlier, see Eq.~(\ref{eq:oscdata}), does not determine the absolute scale or ordering of neutrino masses. The experiments have measured the two mass-squared differences (solar and atmospheric) associated with neutrino oscillations, and the ratio of the solar-to-atmospheric mass-squared difference ($r$) is found to be~\cite{Esteban:2020cvm}
\begin{eqnarray}\label{eq:r}
   r=\Delta m_{21}^2/|\Delta m_{31}^2|\sim 0.03.  
\end{eqnarray}
This may be an indication of the involvement of two different mass scales that might originate from entirely separate mechanisms\footnote{Within the framework of type-I seesaw mechanism, the hierarchy of the atmospheric and solar neutrino mass scales can also be explained through the mechanisms of sequential dominance~\cite{King:1998jw, King:1999cm, King:1999mb, King:2002nf, Antusch:2010tf, Antusch:2004gf} and constrained sequential dominance ~ \cite{King:2005bj, Antusch:2011ic}  where one of the right-handed neutrinos is considered to dominantly contribute the light neutrino masses.}. Following this ethos, the authors in~\cite{Rojas:2018wym} showed a  scotogenic extension of the type-I seesaw scenario that can minimally explain the hierarchy of solar and atmospheric neutrino mass scales.
In this set-up, the Standard Model particle content is extended by including one heavy isosinglet neutral lepton $N_R$ (for the type-I sector) along with a  dark fermion $f$ and an inert scalar doublet $\eta$ (for the scotogenic sector), both being odd under a dark $Z_2$ symmetry. Here, the type-I contribution gives rise to the larger atmospheric scale. In contrast, the one-loop scotogenic contribution turns out to be the origin of the smaller solar mass scale mediated by a dark fermion, also providing a potential dark matter candidate~\cite{Leite:2023gzl, Mandal:2021yph}.   Unfortunately, such constructions turn out to be inadequate in explaining the observed neutrino oscillation data associated with the mixing and the Dirac CP phase mentioned above. This problem can be addressed by augmenting well-motivated non-Abelian discrete flavor symmetries.  
Earlier, in~\cite{Barreiros:2020gxu, Ganguly:2022qxj}, the authors proposed flavor symmetric realization of scoto-seesaw framework with two right-handed neutrinos with $Z_8$ and $A_4$ discrete symmetries respectively. The use of $A_4$  discrete flavor symmetry has the additional advantage of obtaining analytic expressions for neutrino masses and mixing angles as well as yields interesting correlations among the oscillation parameters with distinctive predictions. For instance, in \cite{Ganguly:2022qxj}, the authors realized TM$_2$ mixing scheme within a flavor symmetric scoto-seesaw (FSS) framework with $A_4$ symmetry which can arise in various ways such as starting from a continuous~\cite{Koide:2007sr, Adulpravitchai:2009kd, Luhn:2011ip, Merle:2011vy, Rachlin:2017rvm, King:2018fke} or super-string theory in compactified extra dimensions~\cite{Altarelli:2005yp, Altarelli:2005yx, Burrows:2009pi, King:2017guk, deAdelhartToorop:2011re, Feruglio:2017spp, deAnda:2018ecu, Novichkov:2018yse, Criado:2018thu, Kobayashi:2018wkl, Penedo:2018nmg, Ding:2019xna}. Furthermore, in another $A_4$ based scoto-seesaw framework~\cite{Bonilla:2023pna}, the authors showed the hierarchy of atmospheric and solar neutrino mass splitting can be obtained as a prediction of the `discrete dark matter' mechanism~\cite{Hirsch:2010ru}. In this construction, both scotogenic contribution and the stabilizing symmetry for DM  (obtained as a residual symmetry of $A_4$ breaking) appears naturally, however, additional copies of right-handed neutrino and scalar doublet are essential. 

In the present work, we show that the scoto-seesaw mechanism can be embedded within  $A_4\times Z_4 \times Z_3 \times Z_2$ flavor symmetric framework with the involvement of {\it only one right-handed neutrino}, and the {\it experimentally favored  TM$_1$ mixing scheme} can be realized. We will call the present model FSS$_1$ from now on as it explains the TM$_1$ mixing scheme. The scotogenic contribution contains neutral particles in both fermionic and scalar sectors. Within this flavor symmetric framework, we perform a phenomenological study of the fermionic dark matter and determine its correlation with the observed neutrino mixing. The obtained magnitude and flavor structure of the scotogenic contribution dictates the observed neutrino mixing pattern and facilitates us in obtaining the correct dark matter relic density. Such a direct correlation with the dark matter sector was missing in \cite{Ganguly:2022qxj}, a detailed analysis on fermionic dark matter in this work also completes the associated phenomenology. The presence of flavor symmetry makes an interesting prediction for the neutrino mass hierarchy, determines the octant of the atmospheric mixing angle $\theta_{23}$ and tightly constrains the TM$_1$  prediction for the Dirac CP phase $\delta_{\rm CP}$. Here, the type-I contribution produces a neutrino mass matrix of rank 1, yielding only one massive light neutrino. Subsequently, the scotogenic contribution generates another neutrino mass eigenstate, and together, we obtain two massive neutrinos, which follow the normal ordering of light neutrino mass spectrum. In~\cite{Rojas:2018wym}, the authors showed that the type-I and scotogenic contributions could be the origin of the atmospheric and solar mass scales. {\it Here, we show that such a hierarchy can also be procured within a flavor symmetric construction, explaining observed neutrino oscillation data.} Furthermore, in the present model, we can obtain the constraint on the Majorana phase and predict the effective mass parameter appearing in the neutrinoless double beta decay and the sum of the absolute neutrino masses.

One of the essential aspects of any theoretical model is its experimental viability. For the version of the FSS model discussed here, we perform a comprehensive phenomenological analysis involving the $h\rightarrow \gamma \gamma$ decay, where $h$ is the SM Higgs boson. The signal strength of the Higgs in the diphoton decay channel, $R_{\gamma\gamma}$, is measured at the LHC, the value of which is around one~\cite{ATLAS:2022tnm}. The additional contribution to the decay of $h\rightarrow \gamma\gamma$ in the FSS$_1$ model is the charged scalar of the $\eta$ field. Our analysis shows that $R_{\gamma\gamma}$ can be fitted in our model, which can constrain the mass of the charged component of the $\eta$ field. Owing to the flavor structure of this scoto-seesaw framework, we find that only the scotogenic part contributes to the lepton flavor violating decays such as $\mu \rightarrow e\gamma, \mu \rightarrow 3e$, whereas only the seesaw part contributes in the decays such as $\tau \rightarrow \mu\gamma, \tau \rightarrow 3 \mu$. However,  scotogenic and seesaw parts do not contribute to the $\tau\rightarrow e\gamma$ and $\tau\rightarrow 3 e$ decays, a distinctive feature of the present construction. All these phenomenological analyses for the FSS$_1$ framework serve as crucial tests of the model’s predictions and provide valuable insights into its compatibility with experimental data.  

The rest of the paper is organized as follows. In section \ref{sec:min-scoto-seesaw}, we briefly introduce the minimal scoto-seesaw framework. In section \ref{sec:A4FSS} we present the complete $A_4$ flavor symmetric scoto-seesaw scenario, and in  section~\ref{sec:nu-mixing} we analyze corresponding neutrino masses and mixing.   We mention the low energy scalar potential in section~\ref{scalar-potential}. In section~\ref{sec:DMpheno}, we discuss the detailed phenomenology of fermionic dark matter and further phenomenological implications for the Higgs to the diphoton decay and lepton flavor violation in section~\ref{sec:Collider} and section~\ref{sec:LFVpheno}, respectively.  Then, in section \ref{sec:summary}, we summarize the phenomenological analysis. Finally, in section~\ref{sec:conc}, we present the conclusion and outlook of the FSS$_1$ framework.

\section{Minimal scoto-seesaw model}\label{sec:min-scoto-seesaw}
In this section, we present the minimal scoto-seesaw model which is introduced in \cite{Rojas:2018wym}. The minimal scoto-seesaw model consists of one\footnote{The number of right-handed neutrinos added~\cite{Schechter:1980gr} to the SM is not fixed as they do not carry any anomaly~\cite{King:2015sfk}.} right-handed neutrino $N_R$, one singlet dark fermion $f$ and one extra scalar doublet $\eta_R$. In addition to these particles, one $Z_2$ symmetry is introduced to stabilize the dark matter. In this model\footnote{For various extensions of the minimal scoto-seesaw scenario, see Refs.~{\cite{Barreiros:2020gxu, Mandal:2021yph, Ganguly:2022qxj, Leite:2023gzl, Kumar:2023moh, VanDong:2023xbd,Kumar:2024zfb,Borah:2024gql}.}}, the usual type-I seesaw mechanism with one right-handed neutrino $N_R$ is combined with the scotogenic model with fermion $f$. The type-I seesaw generates the atmospheric mass scale at the tree level, while the solar mass scale is generated at the loop level in the scotogenic mechanism. As a result, the hierarchy between solar mass scale and atmospheric mass scale is maintained. The relevant  Lagrangian in the model can be written as
\begin{eqnarray}\label{eq:scoto-seesaw Lag}
        {\mathcal{L}}=-Y_{N}^k \bar{L}^ki\sigma_2 H^* N_R+\frac{1}{2}M_N \bar{N}_R^cN_R+ Y_{f}^k \bar{L}^ki\sigma_2 \eta^* f +\frac{1}{2}M_f \bar{f}^c f + h.c..
\end{eqnarray}
where $L^k$ are the lepton doublets. The scalars $H$ and $\eta$ are the $SU(2)$ doublets defined in Eq.~(\ref{eq:eta}). $Y_N$ and $Y_f$ are complex $3\times 1$ Yukawa coupling matrices, and $M_{N,f}$ are the mass terms for $N_R$ and $f$. The total neutrino mass reads \cite{Rojas:2018wym}
\begin{eqnarray}\label{eq:scoto-seesaw mass matrix}
    M_{\nu}^{ij}=-\frac{v^2}{M_N}Y_N^i Y_N^j+{\mathcal{F}}(M_{\eta_R},M_{\eta_I},M_f) M_f~Y_{f}^i Y_f^j.
 \end{eqnarray}
Here, the first term is due to the tree-level seesaw mechanism, while the second term originates from the one-loop scotogenic contribution with
\begin{eqnarray}\label{eq: loop function F}
    \mathcal{F}(M_{\eta_R},M_{\eta_I},M_f)=\frac{1}{32 \pi^2}\Big[\frac{M_{\eta_R}^2 \log\Big(M_f^2/M_{\eta_R}^2\Big)}{M_f^2-M_{\eta_R}^2}-\frac{M_{\eta_I}^2 \log\Big(M_f^2/M_{\eta_I}^2\Big)}{M_f^2-M_{\eta_I}^2}\Big],
\end{eqnarray}
where $M_{\eta_R}$ and $M_{\eta_I}$ are the masses of the neutral component of $\eta$.  
Although the ratio of the above two contributions in Eq.~(\ref{eq:scoto-seesaw mass matrix}) can explain the hierarchy of the solar and atmospheric mass scales, it fails to explain the observed neutrino mixing pattern. In this regard, the use of non-Abelian discrete flavor symmetries is well motivated~\cite{Altarelli:2010gt, King:2013eh, Petcov:2017ggy, Chauhan:2022gkz, Chauhan:2023faf}. In the following sections, we discuss the phenomenological consequences of flavor symmetric construction of the scoto-seesaw framework with only one right-handed neutrino to explain the correct neutrino masses and mixing.  We also provide a detailed analysis of the fermionic dark matter relic abundance and direct detection search constraint to determine the parameter space consistent with neutrino oscillation data and predictions for Higgs to diphoton signal strength and lepton flavor violating decays. 

\section{Scoto-seesaw with flavor  symmetry: the FSS$_1$ model}\label{sec:A4FSS}
The model we are proposing is the flavor symmetric version of the scoto-seesaw model described in the previous section with usual scotogenic fermion $f$ and inert doublet $\eta$ in addition to one right-handed neutrino $N_R$. To obtain the flavor structure, $A_4 \times Z_4 \times Z_3 \times Z_2$ flavor discrete symmetry and flavons $\phi_s$, $\phi_a$, $\phi_T$ and $\xi$ are introduced. To avoid unwanted terms in the Lagrangian and get the correct flavor Yukawa structure, additional $Z_N$ symmetries are introduced. The inclusion of flavon fields and auxiliary symmetries are generic features of such flavor symmetric constructions~\cite{Ma:2001dn,Ma:2002yp,Altarelli:2005yp,Altarelli:2005yx}. A remnant $Z_2$ symmetry of the $Z_N$ symmetries acts as a dark symmetry that ensures the stability of dark matter under which only $f$ and $\eta$ are odd. Similar types of flavored scoto-seesaw models were studied before in \cite{Barreiros:2020gxu} and \cite{Ganguly:2022qxj} with $Z_8$ and $A_4 \times Z_4 \times Z_3 \times Z_2$ discrete symmetries, respectively. No simple analytic correlation can be obtained due to the use of the $Z_8$ symmetry~\cite{Barreiros:2020gxu} whereas the TM$_2$ mixing was reproduced in~\cite{Ganguly:2022qxj}  with the $A_4 \times Z_4 \times Z_3 \times Z_2$ symmetry. In both cases, two right-handed neutrinos are introduced in the seesaw contribution to get the flavor structure and mixing. In the present work, we construct the framework with only one right-handed neutrino and realize the experimentally preferred TM$_1$ mixing scheme compared to the TM$_2$ mixing scheme (derived in~\cite{Ganguly:2022qxj}).   The particle content of our model and charge assignment under different symmetries are shown in Table \ref{table:A4 table}. 
The role of each discrete auxiliary symmetry will be described in detail as we proceed further. 
  \begin{table}[h!]
		\begin{center}
			\begin{tabular}{ccccccc|cccc}
				\hline
				Fields & $e_R$, $\mu_R$, $\tau_R$ & $L_{\alpha}$ & $H$ & $N_{R}$&  $f$& $\eta$ & $\phi_s$& $\phi_a$ & $\phi_T$ & $\xi$ \\
				\hline
				$A_4$ & $1$ , $1^{\prime\prime}$ , $1^{\prime}$ & $3$ & $1$ & $1$ & $1$ & $1$ & $3$ &$3$& 3 & $1^{\prime\prime}$ \\
				$Z_4$ &  $-1$ & $i$ & $1$ & $1$ & $1$ & $1$  & $i$ & $i$ & $-i$& $1$ \\
				$Z_3$ &  $1$ & $\omega$ &$\omega$ & $1$& 1 & $1$ & $\omega^2$ & $\omega$ & $1$ & $1$ \\
				$Z_2$ & $-1$  & $1$ & $1$ & $1$ & $-1$ & $-1$ & $1$ & ${-1}$ & $-1$ & ${-1}$ \\
			    \hline
			\end{tabular}
		\caption{ Fields content and transformations under the symmetries of the FSS$_1$ model. Product rules of the $A_4$ singlets and triplets are given in Appendix \ref{section:A4 group}. The flavon fields mentioned in the second block are essential to implement the  $A_4$ symmetry, and  $\omega (=e^{2i\pi/3})$ is the cube root of unity. }
		\label{table:A4 table}
		\end{center}
	\end{table}

With the field content and charges assignment in Table \ref{table:A4 table}, the charged lepton Lagrangian can be written up to leading order as 
\begin{eqnarray}\label{eq:Lag-cl}
	\mathcal{L}_l=\frac{y_e}{\Lambda}(\bar{L}\phi_T)H e_R + \frac{y_{\mu}}{\Lambda}(\bar{L}\phi_T)H \mu_R + \frac{y_{\tau}}{\Lambda}(\bar{L}\phi_T)H \tau_R + h.c., \end{eqnarray}
 where $\Lambda$ is the cut-off scale of our model. $y_e$, $y_{\mu}$ and $y_{\tau}$ are the coupling constants. Now, when the flavon $\phi_T$ gets a vacuum expectation value (VEV) in the direction $\langle \phi_T \rangle=(v_T,0,0)^T$ and subsequently the Higgs field   also get  VEV as $\langle H \rangle=v$, where $v$ is the SM VEV, we get the charged lepton mass matrix to be in the diagonal form as
	\begin{eqnarray}\label{eq:cl-mass-matrix}
	M_l=\frac{v_T}{\Lambda}v\begin{pmatrix}
	y_e & 0 & 0 \\
	0 & y_{\mu} & 0 \\
	0 & 0 & y_{\tau}
	\end{pmatrix}. 
	\end{eqnarray}
Now, the Lagrangian in the neutrino sector, which generates neutrino masses, constitutes two parts: a type-I seesaw contribution with one right-handed neutrino $N_R$ and another, one loop scotogenic part with the presence of the dark fermion $f$ and scalar $\eta$. Following the symmetries and particle content mentioned in table \ref{table:A4 table}, the Lagrangian for the neutrino sector can be written as
\begin{eqnarray}\label{eq:Lag}
	\mathcal{L} = \frac{y_{N}}{\Lambda}(\bar{L}\phi_s)\tilde{H} N_{R}+\frac{1}{2}M_{N} \bar{N}_{R}^c N_{R}+ \frac{y_s}{\Lambda^2}(\bar{L}\phi_a)\xi i\sigma_2 \eta^* f+ \frac{1}{2}M_f \bar{f}^c f+ h.c.,
\end{eqnarray} 
where $y_N$ and $y_s$ are the coupling constants and $M_N$ is the Majorana mass of the right-handed neutrino $N_R$ while $M_f$ is the mass of the fermion $f$. In the above Lagrangian, we have considered VEVs of the flavons $\phi_s$, $\phi_a$ and $\xi$ in directions $\langle \phi_s \rangle=(0,-v_s,v_s)$, $\langle \phi_a \rangle=(2  v_a,v_a,0)$ and  $\langle {\xi} \rangle = v_{\xi}$, respectively. A similar vacuum alignment can be found in the literature for neutrino model building~\cite{Antusch:2011ic,Bjorkeroth:2014vha} which can be realized inherently by analyzing the complete scalar potential \cite{Altarelli:2010gt, Karmakar:2016cvb,  He:2006dk, Lin:2008aj, King:2005bj,Holthausen:2011vd}. The light neutrino mass matrix involving both type-I seesaw  and scotogenic contributions can be written as 
\begin{eqnarray}\label{eq:neutrino-mass-matrix}
 (M_{\nu})_{ij}  &=& -\frac{v^2}{M_N}Y_N^iY_N^j +\mathcal{F}(M_{\eta_R},M_{\eta_I},M_f)M_f Y_f^i Y_f^j
\end{eqnarray}
 where the Yukawa couplings take the following form 
\begin{eqnarray}
   && Y_N=(Y_N^e,Y_N^{\mu},Y_N^{\tau})^T=(0,y_N \frac{v_s}{\Lambda},- y_N \frac{v_s}{\Lambda})^T,\label{eq:seesaw yukawa}\\
    && Y_F=(Y_F^e,Y_F^{\mu},Y_F^{\tau})^T= (y_s \frac{v_{\xi}}{\Lambda}\frac{v_{a}}{\Lambda},y_s \frac{v_{\xi}}{\Lambda}\frac{2 v_{a}}{\Lambda}, 0)^T  \equiv  (\kappa,2\kappa,0)^T.
\label{eq:scoto yukawa}
\end{eqnarray}
Within this setup, the total effective light neutrino mass matrix of Eq.~(\ref{eq:neutrino-mass-matrix}) is the following
\begin{eqnarray}\label{eq:total-mass-matrix}
	M_{\nu}&=& \begin{pmatrix}
	b & 2b  & 0 \\
	2b & -a+ 4b & a \\
	0 & a & -a
	\end{pmatrix} 
 \end{eqnarray}
with
\begin{eqnarray}\label{eq:def-a}
a&=&y_N^2\frac{v^2}{M_N}\frac{v_s^2}{\Lambda^2}, \\
b&=&y_s^2 \frac{v_{\xi}^2}{\Lambda^2}\frac{v_a^2}{\Lambda^2}\mathcal{F}(m_{\eta_R},m_{\eta_I},M_f)M_f \equiv \kappa^2 \mathcal{F}(M_{\eta_R},M_{\eta_I},M_f)M_f, \label{eq:def-a-b}
\end{eqnarray}
where  $\mathcal{F}$ is the loop function defined in Eq.~(\ref{eq: loop function F}). Clearly, from Eq. (\ref{eq:seesaw yukawa}) to Eq. (\ref{eq:def-a-b}), it is evident that the parameters $a$ and $b$ originate from type I-seesaw and scotogenic contributions, respectively. In the next section, we show how these parameters' relative magnitude helps us explain the hierarchy of the atmospheric and solar oscillation mass scales.   

Though the neutrino mass matrix given in Eq.~(\ref{eq:total-mass-matrix}) is obtained through a combination of type-I seesaw and scotogenic mechanisms, there can be additional operators like $LHLH/\Lambda$,  contributing to the light neutrino masses. In our model,  this higher dimensional term is not invariant explicitly under the $Z_4$ symmetry given in Table~\ref{table:A4 table}. Also, terms like $LHLH (\phi_a, \phi_s, \phi_T,\xi)/\Lambda^2$ are disallowed due to the considered discrete $Z_N$ symmetries. For the same $Z_N$ symmetries, the scotogenic contribution ($\bar{L}i\sigma_2 \eta^* f$) is only allowed at $1/\Lambda^2$ with the involvement of flavons $\phi_a$ and $\xi$, which are both odd under the $Z_2$ symmetry along with $f$ and $\eta$. Here, in principle, there could be another contributing term via $\bar{L}i\sigma_2 \eta^* f \phi_a \xi^{\dagger}/\Lambda^2$. This term, however, can be absorbed in the previous contribution through a redefinition of the Yukawa coupling.  Owing to the $A_4$ symmetry, in the charged lepton sector, the leading order contribution appears only at dimension-5. However, for example, there could be a next-to-leading correction at $\mathcal{O}(1/\Lambda^2)$ via $(\bar{L}\phi_s^{\dagger}\phi_a) H\alpha_R/ \Lambda^2$, where $\alpha_R$ is the corresponding right-handed charged lepton. Interestingly, such a contribution is disallowed due to the $Z_4$ symmetry given in Table~\ref{table:A4 table}. As the right-handed Majorana neutrino present in our model is also a singlet under $A_4$ symmetry, any higher-order correction can be absorbed in the leading order contribution to  $M_N$. For the same reason, we can also absorb any higher order contribution to  $M_f$ as it does not affect the flavor structure of our model. 
Finally, the Dirac Yukawa coupling is allowed at dimension-5 as given in Eq.~(\ref{eq:scoto yukawa}). The next-to-leading order contributions at $\mathcal{O}(1/\Lambda^2)$ such  as $(\bar{L}\phi_a^{\dagger} \phi_T) \tilde{H}N_R$ and $(\bar{L}\phi_s^{\dagger} \phi_T) \tilde{H}N_R$,  are not allowed due to the $Z_4$ symmetry.
 
\section{Neutrino masses and mixings in the FSS$_1$ model}\label{sec:nu-mixing}
\begin{figure}[h]
	\begin{center}
		\includegraphics[width=.39\textwidth]{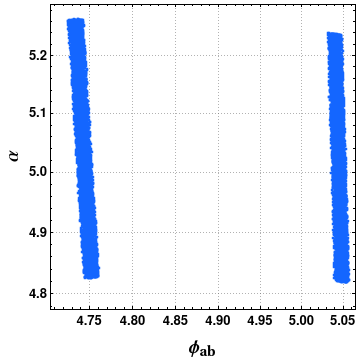}
		\includegraphics[width=.43\textwidth]{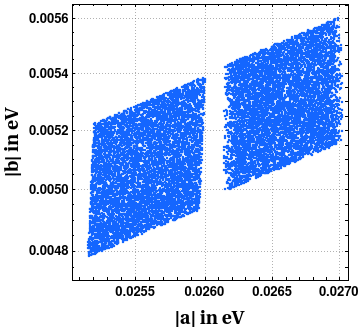}
	\end{center}
\caption{The allowed regions for $\alpha$ and $\phi_{ab}$ (left panel), and $|a|$ and $|b|$ (right panel) for the $3\sigma$ ranges of neutrino oscillation data~\cite{Esteban:2020cvm}.}
\label{fig:alpha}
\end{figure}

The model we presented in the last section has two parts. One is coming from a type-I seesaw with one right-handed neutrino $N_R$. Another part is the scotogenic contribution with the dark fermion $f$. The full light neutrino mass matrix is given in Eq.~ (\ref{eq:total-mass-matrix}), and both contributions are essential in explaining observed neutrino masses and mixing. To diagonalize the mass matrix in Eq.~(\ref{eq:total-mass-matrix}),  we first write the mass matrix in the TBM basis as
\begin{eqnarray}\label{eq:UTB seesaw diag}
     M_{\nu}^{\prime}=U_{TB}^T M_{\nu}U_{TB}=\begin{pmatrix}
    0 & 0 & 0 \\
    0 & 3b & -\sqrt{6}b \\
   0 &  -\sqrt{6}b &  2(b-a)
    \end{pmatrix},
\end{eqnarray}
where
\begin{eqnarray}
    U_{TB}=\begin{pmatrix}
    \sqrt{\frac{2}{3}} & \sqrt{\frac{1}{3}} & 0 \\
    -\sqrt{\frac{1}{6}} & \sqrt{\frac{1}{3}} & -\sqrt{\frac{1}{2}} \\
    -\sqrt{\frac{1}{6}} & \sqrt{\frac{1}{3}} & \sqrt{\frac{1}{2}}
    \end{pmatrix}.
\end{eqnarray}
As evident from Eq. (\ref{eq:UTB seesaw diag}), a further rotation by $U_{23}$ (another unitary matrix) in the 23 plane will diagonalize the light neutrino mass matrix  via $M_{\nu}^{\rm diag}=U_{23}^T M_{\nu}^{\prime}U_{23}$. The unitary rotation matrix $U_{23}$ can be parameterized  as 
\begin{eqnarray}\label{eq:u23}
    U_{23}=\begin{pmatrix}
    1 & 0 & 0 \\
    0 & \cos\theta & \sin\theta e^{-i\psi} \\
    0 & -\sin\theta e^{i\psi} & \cos\theta
    \end{pmatrix}
\end{eqnarray}
where $\theta$ and $\psi$ are the rotation angle and the associated phase factor,  respectively. 
So, the diagonalization of  $M_{\nu}$ can be achieved through
\begin{eqnarray}
    (U_{TB}U_{23})^T M_{\nu}(U_{TB}U_{23})={\rm diag}(m_1 e^{i\gamma_1},m_2 e^{i\gamma_2},m_3 e^{i\gamma_3})
\end{eqnarray}
where $m_{1,2,3}$ are the real and positive mass eigenvalues, and $\gamma_{1,2,3}$ are the phases that are extracted from the corresponding complex eigenvalues. In our framework, we have only one right-handed neutrino, which, via type-I seesaw (the first term in Eq. (\ref{eq:neutrino-mass-matrix})), yields a  rank 1 mass matrix which makes one light neutrino massive. Together with the scotogenic contribution, we obtain a rank 2 mass matrix given in Eq. (\ref{eq:total-mass-matrix}),  generating two massive neutrinos. Hence,  within this flavor symmetric construction,  one mass eigenvalue (lightest) will be zero. So, we have $m_1=0$, which implies $\gamma_1=0$. Now, we can get the form of neutrino mixing matrix $U_{\nu}$ such that $U_{\nu}^T M_{\nu}U_{\nu}={\rm diag}(0,m_2,m_3)$. Thus, $U_{\nu}$ becomes $U_{\nu}=U_{TB}U_{23}U_m$, where $U_m={\rm diag}(1,1,e^{i\alpha_{32}/2})$ is the Majorana phase matrix with $\alpha_{32}=\gamma_3-\gamma_2$. {\it Therefore, we have only one non-zero phase in the Majorana phase matrix $U_m$ as the lightest neutrino is massless.} The explicit form of $U_{\nu}$ follows
\begin{eqnarray}\label{eq:unu}
    U_{\nu}=\begin{pmatrix}
    \sqrt{\frac{2}{3}} & \frac{\cos\theta}{\sqrt{3}} & \frac{e^{-i\psi}\sin\theta}{\sqrt{3}} \\
    -\frac{1}{\sqrt{6}} & \frac{\cos\theta}{\sqrt{3}}+\frac{e^{i\psi}\sin\theta}{\sqrt{2}} & -\frac{\cos\theta}{\sqrt{2}}+\frac{e^{-i\psi}\sin\theta}{\sqrt{3}} \\
    -\frac{1}{\sqrt{6}} &\frac{\cos\theta}{\sqrt{3}}-\frac{e^{i\psi}\sin\theta}{\sqrt{2}} & \frac{\cos\theta}{\sqrt{2}}+\frac{e^{-i\psi}\sin\theta}{\sqrt{3}}
    \end{pmatrix}U_m. 
\end{eqnarray}
This form of $U_{\nu}$ is well known in the literature as a deviation from $U_{\rm TBM}$ and is called the TM$_1$ mixing pattern, where the first column of the lepton mixing matrix is trimaximal. The VEV alignment of the flavons $\phi_{a,s}$ mentioned earlier plays a crucial role in obtaining such mixing pattern.
The lepton mixing matrix $U_{\nu}$ can now be compared with $U_{PMNS}$ which in its standard parametrization is given by~\cite{ParticleDataGroup:2022pth}
\begin{eqnarray}\label{eq:upmns}
U_{\rm PMNS}=\begin{pmatrix}
    c_{12}c_{13} & s_{12}c_{13} & s_{13}e^{-i\delta_{\rm CP}} \\
    -s_{12}c_{23}-c_{12}s_{23}s_{13}e^{i\delta_{\rm CP}} & c_{12}c_{23}-s_{12}s_{23}s_{13}e^{i\delta_{\rm CP}} & s_{23}c_{13} \\
    s_{12}s_{23}-c_{12}c_{23}s_{13}e^{i\delta_{\rm CP}} & -c_{12}s_{23}-s_{12}c_{23}s_{13}e^{i\delta_{\rm CP}} & c_{23}c_{13}
    \end{pmatrix}U_m,
\end{eqnarray}
where $c_{ij}=\cos\theta_{ij}$, $s_{ij}=\sin\theta_{ij}$,  $\delta_{\rm CP}$ is the Dirac CP violating phase and $U_{m}$ is the Majorana  phase matrix. We can see that the total light neutrino mass matrix of Eq. (\ref{eq:total-mass-matrix}) contains two parameters $a$ and $b$ associated with the type-I seesaw and scotogenic contributions, which can be complex in general. We can write these parameters as $a=|a|e^{i\phi_a}$ and $b=|b|e^{i\phi_b}$ where $\phi_a$ and $\phi_b$ are the associated phases. For calculational purpose,  we define the parameter $\alpha=|a|/|b|$ and the difference of phases by $\phi_{ab}=\phi_a-\phi_b$. As $M_{\nu}^{\prime}$ is diagonalized by $U_{23}$, the rotation angle $\theta$ and the phase $\psi$ appearing in Eq. (\ref{eq:u23}) can be expressed in terms of the model parameters as
\begin{eqnarray}\label{eq:theta phi exp}
    \tan\psi=\frac{2\alpha\sin\phi_{ab}}{5-2\alpha\cos\phi_{ab}},\quad \tan2\theta=\frac{2\sqrt{6}}{\cos\psi+2\alpha\cos(\psi+\phi_{ab})}. 
\end{eqnarray}
As the charged lepton mass matrix is diagonal, to obtain the correlation among the mixing angles and phases, we can compare $U_{\nu}=U_{TB}U_{23}U_m$ of Eq.~(\ref{eq:unu}) with $U_{\rm PMNS}$ given in Eq. (\ref{eq:upmns}). These correlations can be written as~\cite{deMedeirosVarzielas:2012apl,Shimizu:2017fgu,Luhn:2013lkn} 
\begin{eqnarray}\label{eq:tm1-corel}
   \sin\theta_{13}e^{-i\delta_{\rm CP}}=\frac{e^{-i\psi}\sin\theta}{\sqrt{3}},\quad \sin^2\theta_{12}=1-\frac{2}{3-\sin^2\theta},~~ \sin^2\theta_{23}=\frac{1}{2}\Big(1-\frac{\sqrt{6}\sin2\theta\cos\psi}{3-\sin^2\theta}\Big). 
\end{eqnarray}
The above relations among the three mixing angles imply a mutual correlation. These correlations are the unique feature of the considered $A_4 \times Z_4 \times Z_3 \times Z_2$ flavor symmetry, giving rise to the  TM$_1$ mixing scheme. More specifically, relations in Eq.~(\ref{eq:tm1-corel}) are general for the TM$_1$ mixing scheme \cite{Luhn:2013lkn,deMedeirosVarzielas:2012apl,Shimizu:2017fgu} where the mixing angles $\theta_{13}$, $\theta_{12}$ are being correlated to each other. The correlation plot among these mixing angles can be found in Ref.~\cite{deMedeirosVarzielas:2012apl} where $\sin^2\theta_{12}$ is restricted to some narrow range corresponding to the $3\sigma$ regions of $\sin^2\theta_{13}$.  Relations in Eq.~(\ref{eq:theta phi exp}) are unique for the considered FSS$_1$ model. From Eqs.~(\ref{eq:theta phi exp})-(\ref{eq:tm1-corel}), it is clear that the angle  $\theta$ and the associated  phase $\psi$ in $U_{23}$ can be linked with the parameters involved in $M_{\nu}$.   Relations in Eq.~(\ref{eq:tm1-corel}) imply that $\delta_{\rm CP}=\psi$ when $\sin\theta>0$, and $\delta_{\rm CP}=\psi \pm \pi$ for $\sin\theta<0$ which can be written in a compact form as $\tan\delta_{\rm CP}=\tan\psi$. Now, from Eq.~(\ref{eq:total-mass-matrix}), the complex mass eigenvalues are calculated to be 
\begin{eqnarray}
    m_1^c&=&0,\label{eq:complex mass eigenvalues1}  \\
    m_2^c&=& \frac{1}{2}\Big(-2a+5b-\sqrt{4 a^2+4ab+25 b^2}\Big),\label{eq:complex mass eigenvalues2}  \\
    m_3^c&=& \frac{1}{2}\Big(-2a+5b+\sqrt{4 a^2+4ab+25 b^2}\Big).\label{eq:complex mass eigenvalues3}
    \end{eqnarray}
    The real and positive mass eigenvalues are calculated as
    \begin{eqnarray}
   m_1&=&0, \label{eq:real mass eigenvalue1} \\
   m_2&=& \frac{|b|}{2}\big[(5-2\alpha\cos\phi_{ab}-P)^2+(Q+2\alpha\sin\phi_{ab})^2\big]^{1/2}, \label{eq:real mass eigenvalue2}\\
   m_3&=&\frac{|b|}{2}\big[(5-2\alpha\cos\phi_{ab}+P)^2+(Q-2\alpha\sin\phi_{ab})^2\big]^{1/2} \label{eq:real mass eigenvalue3}.
    \end{eqnarray}
    where
    \begin{eqnarray}
    &&   P^2=\frac{M\pm \sqrt{M^2+N^2}}{2},\quad Q^2=\frac{-M\pm \sqrt{M^2+N^2}}{2}, \\
    && M=25+4\alpha\cos\phi_{ab}+4\alpha^2\cos2\phi_{ab},\quad N= 4\alpha \sin\phi_{ab}+4\alpha^2\sin2\phi_{ab}.    
    \end{eqnarray}
Now, from Eq.(\ref{eq:complex mass eigenvalues1}) to Eq. (\ref{eq:complex mass eigenvalues3}), we get the phases associated with the complex eigenvalues $m^c_{1,2,3}$. These phases can be written as $\gamma_i=\phi_b+\phi_i$, $i=2,3$. $i=1$ is excluded here as the lightest mass eigenvalue is zero, the phase associated with $m_1^c$ is $\gamma_1=0$. Now, $\phi_{2,3}$ in our model can be written as
    \begin{eqnarray}
        \phi_2=\tan^{-1}\Big(\frac{Q+2\alpha\sin\phi_{ab}}{5-2\alpha\cos\phi_{ab}-P}\Big),\quad \phi_3=\tan^{-1}\Big(\frac{Q-2\alpha\sin\phi_{ab}}{5-2\alpha\cos\phi_{ab}+P}\Big)
    \end{eqnarray}
Using the above relations, we can calculate the  Majorana phase $\alpha_{32}$ in $U_{m}$, which can be written as 
\begin{eqnarray}\label{eq:majorana phase}
    \alpha_{32}=\tan^{-1}\Big(\frac{Q-2\alpha\sin\phi_{ab}}{5-2\alpha\cos\phi_{ab}+P}\Big)-\tan^{-1}\Big(\frac{Q+2\alpha\sin\phi_{ab}}{5-2\alpha\cos\phi_{ab}-P}\Big).
\end{eqnarray}
The phase $\phi_b$ is irrelevant while calculating the Majorana phase as it is the difference between $\gamma_3$ and $\gamma_2$. Finally, the Jarlskog invariant $J_{\rm CP}$ \cite{Jarlskog:1985ht,Krastev:1988yu}
\begin{eqnarray}\label{eq:jcp}
    J_{\rm CP}={\mathcal{I}}(U_{11}U_{22}U_{12}^* U_{21}^*)=s_{12}c_{12}s_{13}c_{13}^2 s_{23}c_{23}\sin\delta_{\rm CP}
\end{eqnarray}
will be used to quantify the $CP$ violation in the FSS$_1$ model.   From Eqs.~(\ref{eq:theta phi exp})-(\ref{eq:majorana phase}), we observe that the mixing angles and all the phases depend on parameters $\alpha$ and $\phi_{ab}$ while the light neutrino masses depend on these parameters as well as on $|b|$. Now, we will estimate these model parameters ($\alpha$, $|b|$ and $\phi_{ab}$ ) using neutrino oscillation data on neutrino mixing angles and mass squared differences. With measured values~\cite{deSalas:2020pgw, Esteban:2020cvm, Capozzi:2021fjo} of mixing angles $\theta_{13}$, $\theta_{12}$ and $\theta_{23}$,  mass-squared differences   $\Delta m_{21}^2$,  $|\Delta m_{31}^2|$ (mentioned in Eq. (\ref{eq:oscdata}), taken from ~\cite{Esteban:2020cvm}) and the ratio $r$ defined in Eq. (\ref{eq:r}),  we first estimate  $\alpha$ and the phase $\phi_{ab}$ using the $3\sigma$ range of neutrino oscillation data. The allowed ranges for $\alpha$ and $\phi_{ab}$ are plotted in the left panel of Fig \ref{fig:alpha} in the $\alpha-\phi_{ab}$ plane.  Here, we find that the allowed ranges of $\alpha$ vary between $4.82-5.27$ whereas two distinct regions of $\phi_{ab}$ are allowed between $4.72-4.76$ and $5.03-5.06$ radian. 
As mentioned earlier, the effective light neutrino mass matrix in the FSS$_1$ model has rank 2 due to the considered $A_4$ symmetry. Hence we obtain two massive light neutrinos as given in  Eqs.~(\ref{eq:real mass eigenvalue1})-(\ref{eq:real mass eigenvalue3}), predicting  {\it only} the normal ordering (NO) of light neutrino masses. To obtain the absolute values of  $m_2$ and $m_3$, we need to find the overall factor $|b|$ appearing in Eqs.~(\ref{eq:real mass eigenvalue2}) and (\ref{eq:real mass eigenvalue3}). Though the factor $|b|$ cancels out while calculating $r$, it can be calculated by fitting solar or atmospheric mass-squared differences after knowing $\alpha$ and $\phi_{ab}$ from the left panel of Fig. \ref{fig:alpha}. After evaluating $|b|$, $|a|$ can be easily estimated using the relation  $|a|=\alpha|b|$. Hence, in the right panel of Fig. \ref{fig:alpha}, we have plotted the allowed region in the   $|a|-|b|$ plane for the $3\sigma$ range of neutrino oscillation data. Corresponding to two distinct regions of $\phi_{ab}$ in the left panel there also exist two distinct regions of the parameters $|a|$ as shown in the right panel of Fig. \ref{fig:alpha}. Now, from Eqs. (\ref{eq:real mass eigenvalue2}) and (\ref{eq:real mass eigenvalue3}), we find that the light neutrino masses are functions of both $a$ and  $b$,  whose origin lies in the type-I seesaw and the scotogenic contributions, respectively. Since $m_1=0$ in the FSS$_1$ framework,   $m_2$ and $m_3$ are proportional to the solar and atmospheric mass-squared differences. 
\begin{figure}[h]
    \centering
\includegraphics[scale=0.45]{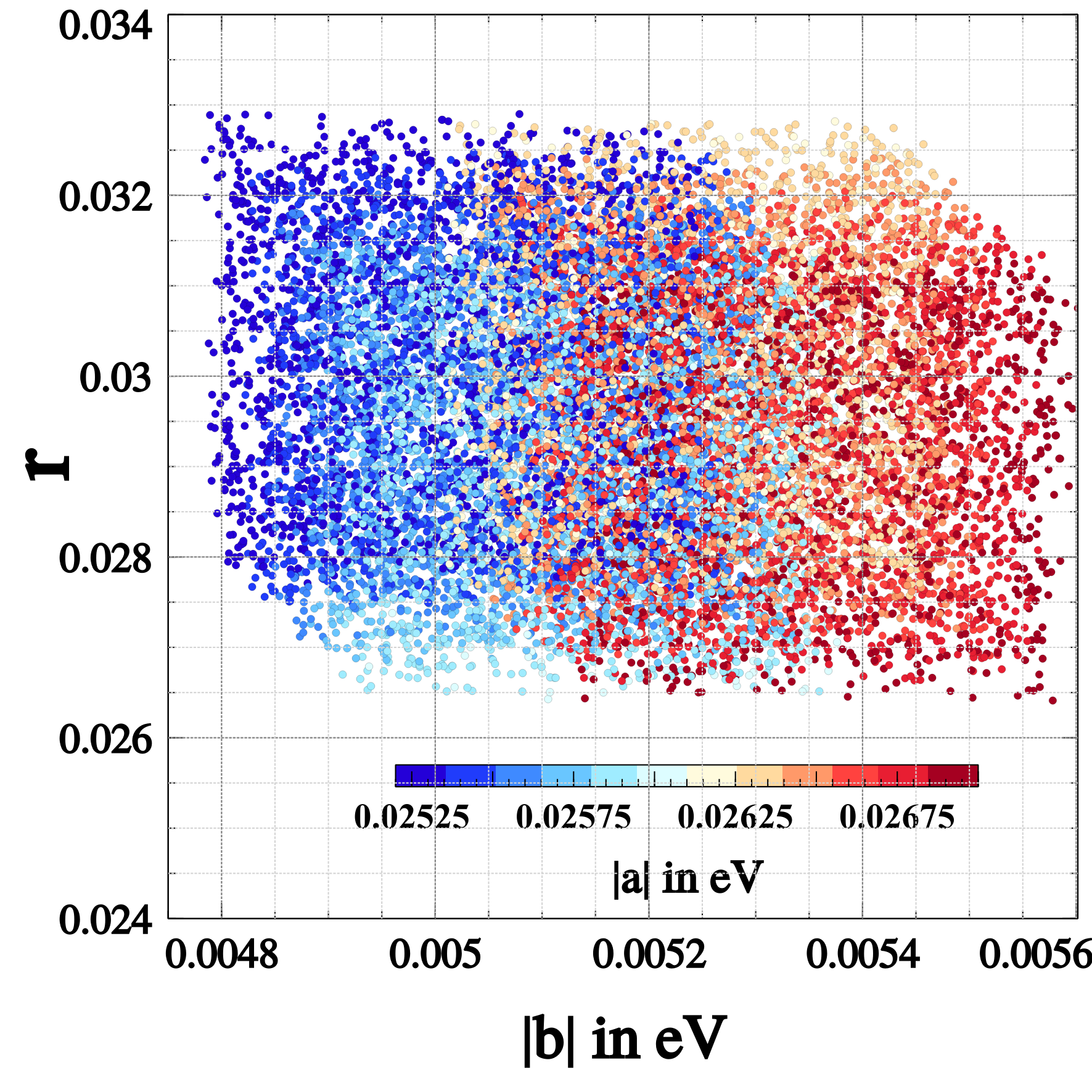}
    \caption{$r$ vs $|b|$ with variation of $|a|$. }
    \label{fig:rvsbla}
\end{figure}
Hence, in Fig~\ref{fig:rvsbla}, we have plotted variation of $|b|$ with respect to $|a|$ (represented by the color variation from blue to red) to reproduce correct $r$. This plot shows that the hierarchy between $|a|$ and $|b|$ essentially explains the observed value of the ratio of the solar to atmospheric mass-squared differences $r$, where $|a|$ is the dominant contribution originated from the type-I seesaw. 
\begin{figure}[h]
	\begin{center}
		\includegraphics[width=.43\textwidth]{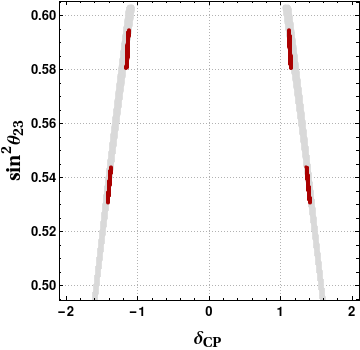}
  \includegraphics[width=.43\textwidth]{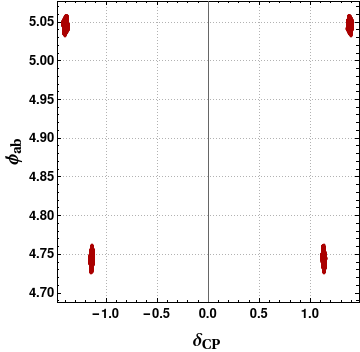}
	\end{center}
\caption{$\sin^2\theta_{23}-\delta_{\rm CP}$ correlation plot (denoted by red dots) is presented in the left panel. The grey-shaded region represents the generic prediction for the TM$_1$ mixing scheme. In the right panel,   $\delta_{\rm CP} $ is plotted against the relative (between $a$ and $b$) phase factor $\phi_{ab}$.  
} 
\label{fig:ss23-dcp}
\end{figure}

With the allowed values of $\alpha$, $\phi_{ab}$, $|a|$ and  $|b|$  obtained from Fig.~\ref{fig:alpha}, we are in a position to study the correlations among neutrino mixing angles, phases, and masses. Due to the presence of the $A_4$ discrete flavor symmetry we have realized the TM$_1$ mixing scheme yielding interesting correlations among the observables appearing in the neutrino mixing. It is well known that there are still some unsettled issues in the measurement of $\theta_{23}$ and $\delta_{\rm CP}$~\cite{deSalas:2020pgw, Esteban:2020cvm, Capozzi:2021fjo}. These are (i) octant of $\theta_{23}$ ( i.e., whether $\theta_{23}<45^o$ or $\theta_{23}>45^o$) and (ii) the precise measurement of $\delta_{\rm CP}$. Following Eqs. (\ref{eq:theta phi exp}) and (\ref{eq:tm1-corel}), we find a correlation between the atmospheric mixing angle $\theta_{23}$ and the Dirac CP phase $\delta_{\rm CP}$ for TM$_1$ mixing scheme. Together with Eq.~(\ref{eq:theta phi exp}) and Fig.~\ref{fig:alpha} within the FSS$_1$ framework, the predictions regarding $\theta_{23}$ and $\delta_{\rm CP}$ for the TM$_1$ scheme gets constrained further as plotted in the left panel of Fig.~\ref{fig:ss23-dcp}.  Here, the gray-shaded region represents the TM$_1$ prediction in the  $\theta_{23}-\delta_{\rm CP}$ plane where the red-shaded region is the prediction for the FSS$_1$ framework. {\it We find that our model prefers the higher octant of $\theta_{23}$ for narrow regions of $\delta_{\rm CP}$.} The allowed regions of $\sin^2\theta_{23}$ are $0.531-0.544$ and $0.580-0.595$ whereas the allowed regions of $\delta_{\rm CP}$ are $\pm(1.44-1.12)$ rad. Here, the relative phase between type-I and scotogenic contributions (denoted by $\phi_{ab}$) is the source of CP violation in the lepton sector, see Eq. (\ref{eq:theta phi exp}) and subsequent discussion. Hence, in the right panel of Fig. \ref{fig:ss23-dcp}, we have plotted the dependence of $\delta_{\rm CP}$ on $\phi_{ab}$ (the relative phase between $a$ and $b$) denoted by the red shaded regions. 
\begin{figure}[h]
	\begin{center}
       \includegraphics[width=.43\textwidth]{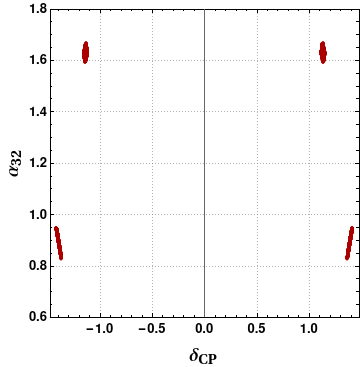}
              \includegraphics[width=.43\textwidth]{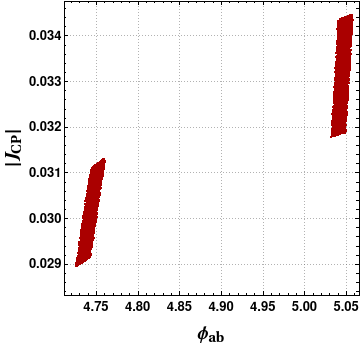}
	\end{center}
\caption{Correlation plot between the Dirac CP phase $\delta_{CP}$ and the Majorana phase $\alpha_{32}$  (left panel) \& Dependence of the Jarlskog invariant $J_{\rm CP}$ on $\phi_{ab}$  (right panel).}
\label{fig:majorana phase}
\end{figure}
It is established that the Majorana phases cannot be constrained from neutrino oscillation data directly,  as they do not appear in the neutrino oscillation probability~\cite{Bilenky:1980cx,Langacker:1986jv,Gluza:2001de}. {\it In the FSS$_1$ framework, the Majorana phase $\alpha_{23}$ can be constrained} using Eq. (\ref{eq:majorana phase}) with the allowed range for $\alpha$ and $\phi_{ab}$. Hence, in the left panel of Fig.~\ref{fig:majorana phase}, we showed the correlation among the CP phases in   $\alpha_{23}-\delta_{CP}$  plane, and the Majorana phase is found to be within the range $0.82-0.95$ and $1.58-1.67$ radian. Estimating the Majorana phase will play a crucial role in predicting the effective mass parameter appearing in the neutrinoless double decay~\cite{Bilenky:1980cx}.   Now, following Eq. (\ref{eq:jcp}), we have plotted the  Jarlskog invariant $J_{\rm CP}$ as a function of $\phi_{ab}$ in the right panel of Fig.~\ref{fig:majorana phase}. Here the magnitude of $J_{\rm CP}$ is found to be within the range $0.0290-0.0313$ and $0.0318-0.0344$. Finally, with the allowed parameter space obtained in Fig. \ref{fig:alpha}, we make predictions for the light neutrino masses ($m_2,m_3$), their sum ($\sum m_i$) and the effective mass parameter appearing in the neutrinoless double decay ($m_{\beta\beta}$) as summarized in Table \ref{tab:mass}. 
\begin{table}
    \centering
    \begin{tabular}{|c|c|c|c|}
        \hline
        $m_2$   & $m_3$  & $\sum m_i$  & $m_{\beta\beta}$  \\
        \hline
        $8.3-9.0$ & $49.7-51.3$ & $58.0-60.3$ & ${1.61-3.85}$\\
        \hline
    \end{tabular}
    \caption{Predictions for $m_2$, $m_3$, $\sum m_i$ and $m_{\beta\beta}$ (all in meV) for the FSS$_1$ framework.}
    \label{tab:mass}
\end{table}
The prediction for $\sum m_i$ is consistent with cosmological observation~\cite{Planck:2018vyg} whereas prediction for $m_{\beta\beta}$ falls below the upper limit provided by the next generation double beta decay experiment nEXO~\cite{nEXO:2021ujk}.


\section{Scalar Potential}\label{scalar-potential}
The FSS$_1$ model considered here consists of two $SU(2)$ doublet scalars $H$ and $\eta$. To obtain the flavor structure of the leptons we have four flavons $\phi_s,\phi_a, \phi_T$ and $\xi$ as mentioned in Tab~\ref{table:A4 table}. These  $SU(2)$ singlet flavons are considered to be very heavy compared to $H$ and $\eta$ and hence remain decoupled from the low energy phenomenology of scalars. The low energy scalar potential of the model can be written as
\begin{eqnarray}
    V&=&-\mu_{1}^2(H^{\dagger}H)+\mu_{2}^2(\eta^{\dagger}\eta)+\lambda_1 (H^{\dagger}H)^2+\lambda_2 (\eta^{\dagger}\eta)^2+\lambda_3 (H^{\dagger}H)(\eta^{\dagger}\eta)+\lambda_4 (H^{\dagger}\eta)(\eta^{\dagger}H)\nonumber \\
    &&+\frac{\lambda_5}{2}\{(H^{\dagger}\eta)(H^{\dagger}\eta)+h.c.\} \label{eq:scalarpot}
\end{eqnarray}
The doublets in the model can be parameterized as
\begin{eqnarray}
    H=\begin{pmatrix}
        H^+ \\
      v/\sqrt{2}+(h+i\zeta)/\sqrt{2}
    \end{pmatrix},  \quad
    \eta=\begin{pmatrix}
        \eta^+ \\
        (\eta_R+i\eta_I)/\sqrt{2} \label{eq:eta}
    \end{pmatrix}.
\end{eqnarray}
The electroweak gauge symmetry is given by
\begin{eqnarray}
    H=\begin{pmatrix}
        0\\
        v/\sqrt{2}
    \end{pmatrix},\quad 
    \eta=\begin{pmatrix}
        0 \\
        0
    \end{pmatrix}.
\end{eqnarray}
The above symmetry breaking pattern ensures that the $Z_2$ symmetry will remain unbroken and results in two CP even scalars ($h,\eta_R$), one CP odd neutral scalar $\eta_I$ in addition to a pair of charged scalars ($\eta^{\pm}$). Due to the dark $Z_2$ symmetry, there is no mixing between $h$ and $\eta_R$, and $h$ plays the role of the SM Higgs boson. The $Z_2$ symmetry also ensures the stability of the lightest scalar ($\eta_R$ or $\eta_I$) that can act as a dark matter candidate. The masses of all scalars can be written in terms of the following parameters
\begin{eqnarray}
    \{\mu_2,\lambda_1,\lambda_2,\lambda_3,\lambda_4,\lambda_5\}.
\end{eqnarray}
These parameters can be written in terms of physical masses of scalars as~\cite{Arhrib:2012ia}
\begin{eqnarray}
 &&   \lambda_1=\frac{m_h^2}{2 v^2},\quad \lambda_3=\frac{2}{v^2}(M_{\eta^\pm}^2-\mu_2^2),\\
&& \lambda_4=\frac{M_{\eta_R}^2+M_{\eta_I}^2- 2 M_{\eta^{\pm}}^2}{v^2},\quad \lambda_5=\frac{M_{\eta_R}^2-M_{\eta_I}^2}{v^2}.
\label{eq:scalarpot1}
\end{eqnarray}
We can choose all the $\lambda$s as free parameters or, equivalently the four physical scalar masses, $\lambda_2$ and $\mu_2$, namely
\begin{eqnarray}
   \{ \mu_2^2,m_h, M_{\eta_R}, M_{\eta_I}, M_{\eta^{\pm}},\lambda_2\}.
\end{eqnarray}
The quartic couplings are constrained theoretically by perturbativity and vacuum stability. We force the scalar potential to be perturbative which requires all quartic couplings of the scalar potential to obey 
\begin{eqnarray}
    |\lambda_i|\leq 8\pi.  \label{eq:8pi}
\end{eqnarray}
To get the scalar potential to be bounded from below, the following conditions can be obtained~\cite{Lindner:2016kqk,Branco:2011iw}
\begin{eqnarray}
    \lambda_{1,2}>0 \quad \text{and} \quad \lambda_3+\lambda_4-|\lambda_5|+2\sqrt{\lambda_1 \lambda_2} >0\quad \text{and} \quad \lambda_3+2\sqrt{\lambda_1 \lambda_2}>0. \label{eq:perturb}
\end{eqnarray}
Eq.~(\ref{eq:perturb}) give constraints based on the bare couplings of the Lagrangian. Another approach with running parameters of the model evaluated at the cut-off scale $\Lambda$ of the theory is possible, see for instance \cite{Mandal:2021yph,  Lindner:2016kqk, Branco:2011iw, Hernandez:2021tii, CarcamoHernandez:2023wzf}. Apart from these theoretical constraints, $\lambda_3$, $\lambda_4$ and $\lambda_5$ given in Eq.~(\ref{eq:scalarpot1}) can also be constrained from experimental and phenomenological constraints. As we will discuss in the subsequent sections,  $\lambda_5$ is crucially relevant in determining the scotogenic Yukawa coupling and hence is constrained from DM relic density, direct search constraints as well as the neutrino phenomenology. Similarly, $\lambda_{3,4}$ can also be constrained from DM direct search as well as SM Higgs diphoton signal strength.

The presence of the doublet scalar $\eta$ in our model can have important consequences in the context of CDF-II $W$-boson mass anomaly~\cite{CDF:2022hxs}, for instance see  \cite{Babu:2022pdn, Fan:2022dck, Asadi:2022xiy, Strumia:2022qkt, Arcadi:2022dmt, Borah:2022zim, Borah:2022obi, Borah:2023hqw},  as it can affect the EW precision observables $S, T$, and $U$ \cite{ParticleDataGroup:2022pth}.  Through the self-energy correction of the W-boson with the doublet scalar in the loop, the $W$-boson mass can be increased from the SM prediction to the value obtained by the CDF-II collaboration. Parameterizing the new physics effects in terms of the $S,T,U$ parameters as~\cite{Zhang:2006vt,CentellesChulia:2022vpz}:
\begin{eqnarray}
    S&=&\frac{1}{12\pi}\ln\frac{M_{\eta^0}^2}{M_{\eta^+}^2},\nonumber \\
    T&=& \frac{G_F}{4\sqrt{2}\pi^2 \alpha_{em}}\Bigg(\frac{M_{\eta^0}^2+M_{\eta^+}^2}{2}-\frac{M_{\eta^0}^2 M_{\eta^+}^2}{M_{\eta^0}^2-M_{\eta^+}^2}\ln\frac{M_{\eta^+}^2}{M_{\eta^0}^2}\Bigg), \\
    U&=& \frac{1}{12\pi}\Bigg[\frac{(M_{\eta^0}^2+M_{\eta^+}^2)(M_{\eta^0}^4-4 M_{\eta^0}^2 M_{\eta^+}^2+M_{\eta^+}^4)\ln\big(\frac{M_{\eta^+}^2}{M_{\eta^0}^2}\big)}{(M_{\eta^+}^2-M_{\eta^0}^2)^3}-\frac{5 M_{\eta^0}^4-22 M_{\eta^0}^2 M_{\eta^+}^2+5 M_{\eta^+}^4}{3 (M_{\eta^+}^2-M_{\eta^0}^2)}\Bigg], \nonumber
    \end{eqnarray}
where $\eta_{0}= (\eta_R+i\eta_I)/\sqrt{2}$, the W boson mass can be written as \cite{Peskin:1991sw}
    \begin{eqnarray}
        M_{W}^2=({M_{W}^2})_{\rm SM}+\frac{\alpha_{em}\cos^2\theta_{W}}{\cos^2\theta_W-\sin^2\theta_W}M_{Z}^2\Bigg[-\frac{1}{2}S+\cos^2\theta_W T+\frac{(\cos^2\theta_W-\sin^2\theta_W)}{4 \sin^2\theta_W}U\Bigg].
    \end{eqnarray}
where $\alpha_{em}$ is the fine structure constant, $\theta_{W}$ is the Weinberg angle, and $({M_{W}^2})_{\rm SM}$ is the SM predicted value of W boson mass. The dominant correction to $M_W$ comes from the $T$-parameter which is very much sensitive to the mass difference between the charged scalar and the neutral scalar components of the inert doublet. And the CDF-II $W$ mass can be obtained if the mass difference between $\eta^+$ and $\eta_0$ is around 80 to 100 GeV. However, we should stress that CDF-II data on $W$ mass are in contradiction with global electroweak $e^+e^-$ fits and recent ATLAS LHC analysis, with systematic uncertainty
improved by 15\% \cite{ATLAS:2023fsi} and optimised reconstruction of the
W-boson transverse momentum \cite{ATLAS:2023llf}.

\section{DM Phenomenology for the FSS$_1$ model}\label{sec:DMpheno}

In this FSS$_1$ framework, both type-I seesaw and scotogenic mechanisms are combined to obtain correct neutrino masses and mixing.  The scotogenic contribution contains two potential candidates for DM: the lightest neutral scalar and the singlet fermion. Determining the DM relic density hinges on these candidates' production mechanisms during the early Universe. While the literature extensively covers the scalar DM phenomenology\footnote{For scalar dark matter phenomenology within the scoto-seesaw framework, see Ref.~\cite{Mandal:2021yph}}, which aligns with the inert doublet model (IDM) perspective, our focus here is on the singlet fermion denoted as $f$, an odd $Z_2$ particle in the scoto-seesaw scenario. We explore various mechanisms that can yield the correct relic density and delve into the associated parameter space. Since $f$ is a gauge singlet, its production mechanism is intricately tied to its Yukawa couplings, see Eq.~(\ref{eq:scoto yukawa}) and Eq.~(\ref{eq:def-a-b}), with SM leptons and the inert doublet scalar $\eta$. The magnitude of these Yukawa couplings plays a pivotal role in determining whether the correct relic density can be achieved through thermal freeze-out or freeze-in mechanism. 

\begin{figure}[h]
    \centering
    \includegraphics[scale=0.5]{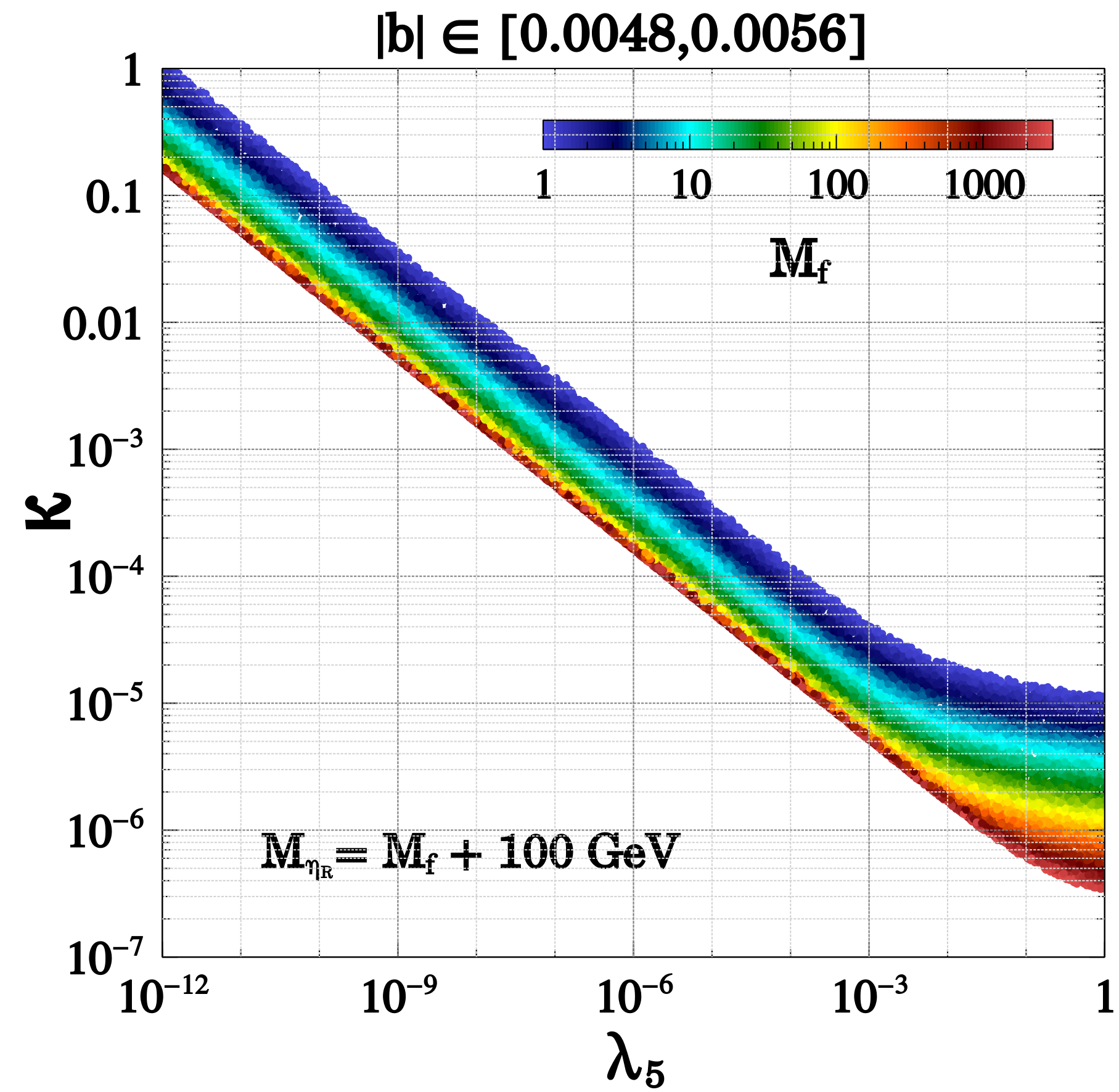}
    \caption{Correlation between $\lambda_5$ defined in Eqs.~(\ref{eq:scalarpot}) and (\ref{eq:scalarpot1}), and $\kappa$ defined in Eq.~(\ref{eq:def-a-b}), consistent with the constrain from neutrino oscillation data, {i.e.} $|b|\in[0.0048,0.0056] $ eV (see Fig.~\ref{fig:alpha}).}
    \label{fig:lam5_kappa}
\end{figure}

\noindent
\underline{Relic Density of DM:}\\ 
\noindent
As outlined in the preceding section, in our FSS$_1$ model, the scotogenic contribution to neutrino mass is parameterized in terms of the parameter $b$ given in Eq.(\ref{eq:def-a-b}), which also plays a crucial role in explaining the observed neutrino oscillation data, is constrained in a range $|b| \in [0.0048,0.0056]$ eV. Thus to obtain the magnitude of Yukawa couplings that can satisfy this constraint with the masses of the loop particles with masses of the order $\mathcal{O}(1-10^{3})$ GeV, we perform a numerical scan, the result of which is shown in the plane of $\kappa$ versus $\lambda_5$ in Fig.~\ref{fig:lam5_kappa}. We have used Eq.~(\ref{eq:def-a-b}) and Eq.~(\ref{eq:scalarpot1}) to obtain the estimations of $\kappa$ and $\lambda_5$. It is worth noticing from the neutrino mass expression that in the limit $\lambda_5\to0$, the scotogenic contribution to neutrino mass vanishes. This is due to the fact that in this limit, the CP-even and CP-odd scalars $\eta_R$ and $\eta_I$ become degenerate and thus $\mathcal{F}(M_{\eta_R},M_{\eta_I},M_f)\to 0$. Thus to satisfy the constraints on $b$, from neutrino oscillation data, if $\lambda_5$ is made small, then $\kappa$ can be enhanced and vice versa. We see that with this constraint on $b$, it is not possible to obtain Yukawa couplings smaller than $\mathcal{O}(10^{-6})$ even if $\lambda_5$ is $\mathcal{O}(1)$. 
Consequently, the singlet fermion also remains in thermal equilibrium with the SM bath. This equilibrium is guaranteed by the doublet scalar $\eta$, which, due to its gauge interactions, consistently maintains equilibrium with the SM bath during the early stages of the Universe. Hence the DM relic density is governed through the WIMP mechanism.
Several pertinent processes contribute to the  relic density of DM. Specifically, the essential parameters influencing the relic density are the Yukawa couplings and the mass differences between the singlet fermion $f$ and other particles in the dark sector, namely $\eta_{R,I},\eta^{\pm}$.

\begin{figure}[h]
	\centering
\includegraphics[scale=0.15]{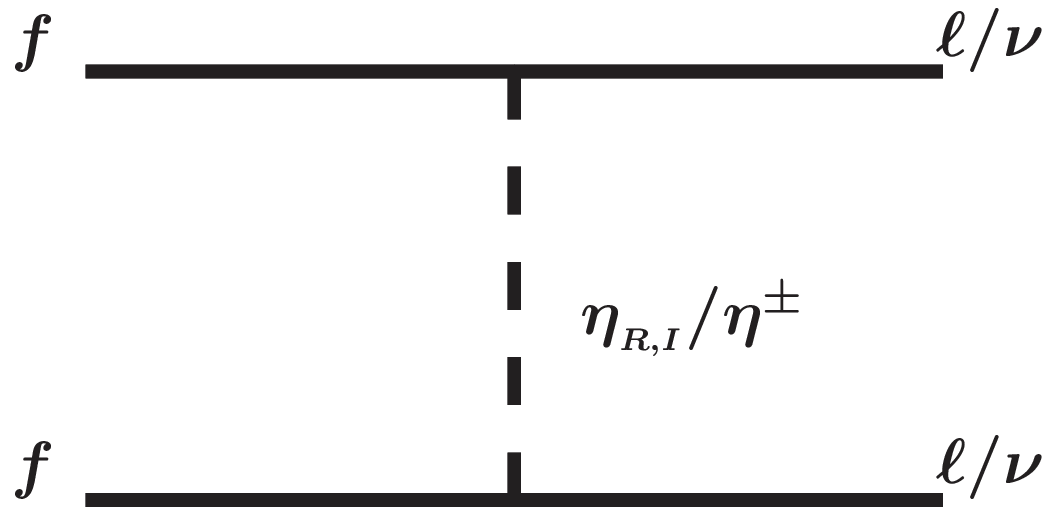}
\caption{Annihilations of dark matter into SM leptons}
\label{fig:dmannfeyn}
\end{figure}

\begin{figure}[h]
	\centering
	\includegraphics[scale=0.3]{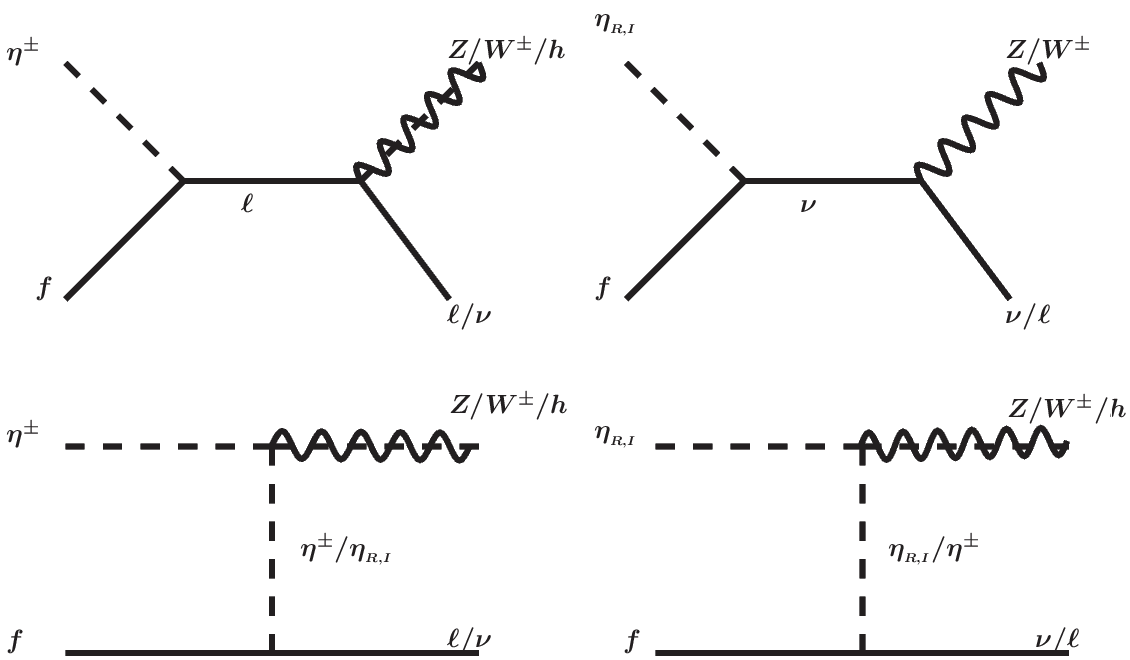}
	\caption{ Co-annihilations among the dark matter and the other dark sector particles.}
 \label{fig:dmcoannfeyn}
\end{figure}

\begin{figure}[h]
	\centering
	\includegraphics[scale=0.35]{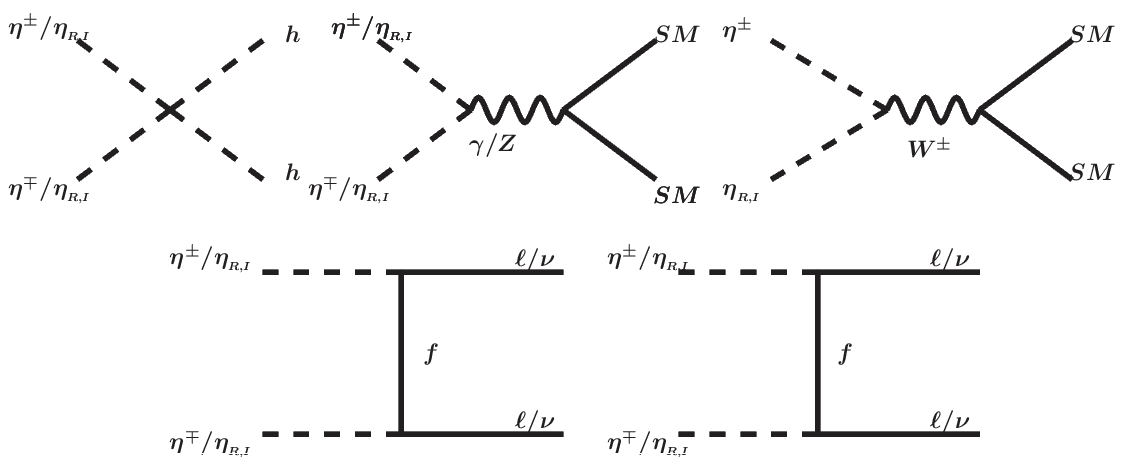}
	\caption{Annihilations of the scalar co-annihilation partners.}
 \label{fig:scalarcoannfeyn}
\end{figure}

For WIMP type DM which is produced thermally in the early universe, its thermal relic abundance can be obtained by solving the Boltzmann equation for the evolution of the DM number density 
\begin{equation}
    \frac{d n}{dt} + 3H n = -\langle \sigma v \rangle_{\rm eff} (n^2 - (n^{eq})^2) 
\end{equation}
where $n=\sum_i n_i $ represents the total number density of all the dark sector particles and  $n^{\rm eq}$ is the equilibrium number density.  $\langle \sigma v \rangle_{\rm eff}$ represents the effective annihilation cross-section which takes into account all number
changing processes for DM freeze-out. It can be written as
\cite{Griest:1990kh}:
\begin{eqnarray}
	\label{eq:effcrs}
	\langle \sigma v \rangle_{\rm eff} &=& \frac{g^2_f}{g^2_{\rm eff}}\langle\sigma v\rangle_{ff}+\frac{g_f g_{\eta_R}}{g^2_{\rm eff}}\langle\sigma v\rangle_{f\eta_R} (1+\Delta_{\eta_R})^{{3}/{2}}\exp(-x\Delta_{\eta_R}) \nonumber \\
 &+& \frac{g_f g_{\eta_I}}{g^2_{\rm eff}}\langle\sigma v\rangle_{f\eta_I} (1+\Delta_{\eta_I})^{{3}/{2}}\exp(-x\Delta_{\eta_I}) \nonumber\\
 &+&\frac{g_f g_{\eta^{\pm}}}{g^2_{\rm eff}}\langle\sigma v\rangle_{f \eta^{\pm}} (1+\Delta_{\eta^{\pm}})^{{3}/{2}}\exp(-x\Delta_{\eta^{\pm}})+\frac{g^2_{\eta_R}}{g^2_{\rm eff}}\langle\sigma v\rangle_{\eta_R \eta_R} (1+\Delta_{\eta_R})^{3}\exp(-2x\Delta_{\eta_R})\nonumber\\&+&\frac{g_{\eta_R}g_{\eta_I}}{g^2_{\rm eff}}\langle\sigma v\rangle_{\eta_R \eta_I} (1+\Delta_{\eta_R})^{3/2}(1+\Delta_{\eta_I})^{3/2}\exp\left(-x(\Delta_{\eta_R}+\Delta_{\eta_I})\right)\nonumber\\&+& \frac{g_{\eta_R} g_{\eta^{\pm}}}{g^2_{\rm eff}}\langle\sigma v\rangle_{\eta_R \eta^{\pm}} (1+\Delta_{\eta_R})^{3/2}(1+\Delta_{\eta^{\pm}})^{3/2}\exp\left(-x(\Delta_{\eta_R}+\Delta_{\eta^{\pm}})\right)\nonumber\\&+&\frac{g^2_{\eta_I}}{g^2_{\rm eff}}\langle\sigma v\rangle_{\eta_I\eta_I} (1+\Delta_{\eta_I})^{3}\exp(-2x\Delta_{\eta_I})+\frac{g^2_{\eta^{\pm}}}{g^2_{\rm eff}}\langle\sigma v\rangle_{\eta^{\pm}\eta^{\mp}} (1+\Delta_{\eta^{\pm}})^{3}\exp(-2x\Delta_{\eta^{\pm}}) \nonumber\\&+&\frac{g_{\eta_I} g_{\eta^{\pm}}}{g^2_{\rm eff}}\langle\sigma v\rangle_{\eta_I \eta^{\pm}} (1+\Delta_{\eta_I})^{3/2}(1+\Delta_{\eta^{\pm}})^{3/2}\exp(-x(\Delta_{\eta_I}+\Delta_{\eta^{\pm}})),
\end{eqnarray} 
where $g_f, g_{\eta_R}, g_{\eta_I} ~{\rm and}~g_{\eta^{\pm}}$ represent the internal degrees of $f,\eta_R,\eta_I$ and $\eta^{\pm}$ respectively and $\Delta_i$ stands for the ratio $(M_i-M_{f})/M_{f}$ with $M_i$ denoting masses of $\eta_R,\eta_{I},\eta^{\pm}$. Here $g_{\rm eff}$ is the effective degree of freedom which is given by
\begin{eqnarray}
	g_{\rm eff}&=& g_f + g_{\eta_R} (1+\Delta_{\eta_R})^{3/2}\exp(-x\Delta_{\eta_R})+ g_{\eta_I} (1+\Delta_{\eta_I})^{3/2}\exp(-x\Delta_{\eta_I})\nonumber\\&+& g_{\eta^{\pm}} (1+\Delta_{\eta^{\pm}})^{3/2}\exp(-x\Delta_{\eta^{\pm}}),
\end{eqnarray}
and $x$ is the dimensionless parameter $M_{f}/T$. 
The relic density of DM $f$ can then be evaluated as :
\begin{eqnarray}
	\Omega_{f} h^2 = \frac{1.09 \times 10^9 {\rm GeV}^{-1}}{\sqrt{g_* }M_{Pl}} \left[\int_{x_{\rm F.O.}}^\infty dx~\frac{\langle \sigma v \rangle_{\rm eff}}{x^2}\right]^{-1}.
\end{eqnarray}
Here $M_{Pl}$ is the Planck mass, $x_{\rm F.O.} =M_{f}/T_{\rm F.O.}$, and $T_{\rm F.O.}$ denotes the freeze-out temperature of $f$.
For this scenario we have implemented the model
in \texttt{micrOMEGAs} \cite{Belanger:2013oya} to calculate the relic abundance of $f$. As evident from Eq.(\ref{eq:effcrs}),  the mass difference between the dark sector particles, namely $f$ and $\eta$, along with the coupling $\kappa$ is pivotal in determining the ultimate relic abundance of dark matter in this configuration. Smaller mass splittings can induce effective co-annihilations between $\eta$ and $f$, potentially reducing the relic abundance to the observed ballpark. The dominant number changing processes relevant in governing the relic density are as shown in Figs.~\ref{fig:dmannfeyn}, \ref{fig:dmcoannfeyn}, and \ref{fig:scalarcoannfeyn}.


Clearly, the processes pivotal in establishing the relic abundance of dark matter fall into three distinct categories: I) the annihilation of dark matter particles into both charged and neutral SM leptons (Fig.~\ref{fig:dmannfeyn}), II) the co-annihilation of dark matter particles with scalar particles from the dark sector (Fig.~\ref{fig:dmcoannfeyn}), and III) the co-annihilation contribution arising from the annihilation of dark-sector scalars (Fig.~\ref{fig:scalarcoannfeyn}). As denoted by Eq.~( \ref{eq:effcrs}), the co-annihilation contribution to the effective annihilation cross-section $\langle \sigma v\rangle_{\rm eff}$ is predominantly shaped by the mass difference between dark matter and the dark scalars. 

\begin{figure}[h]
    \centering
\includegraphics[scale=0.45]{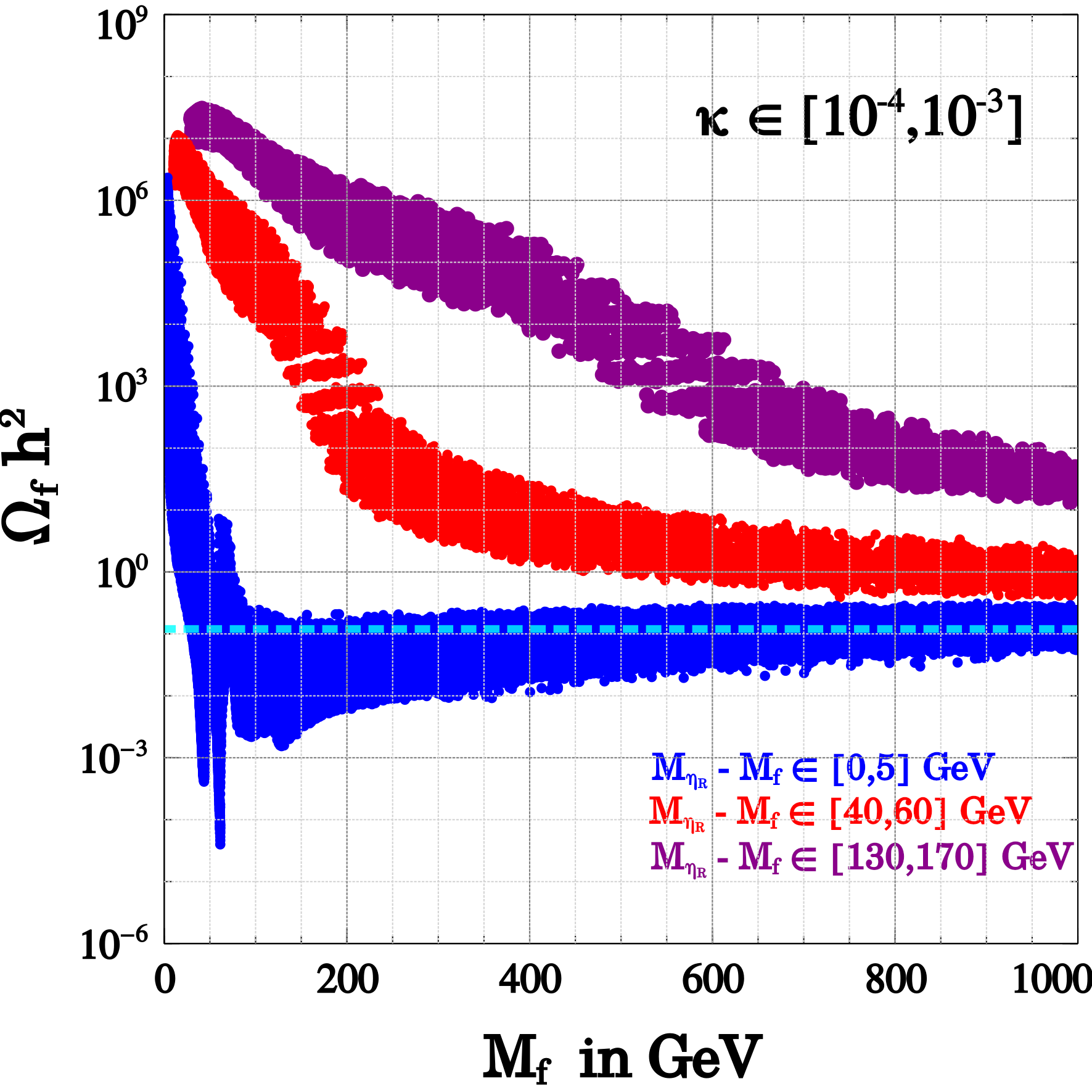}
    \hfil \includegraphics[scale=0.45]{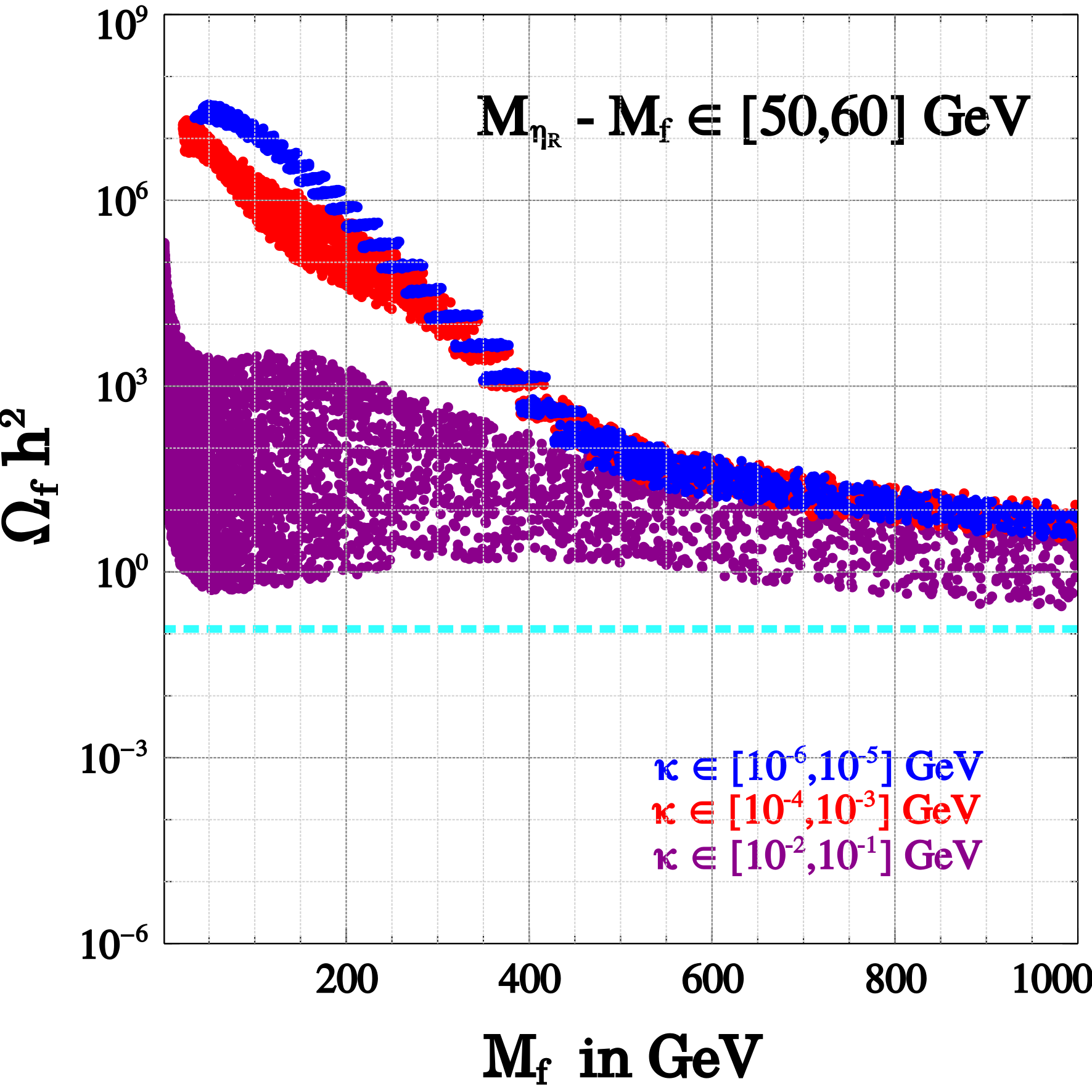}
    \caption{DM relic density as a function of DM mass with the Yukawa couplings (left panel) and mass difference $M_{\eta_R}-M_f$ (right panel) varied randomly as mentioned in the inset of the figure. Horizontal cyan line represents observed relic density~\cite{Aghanim:2018eyx}}.
    \label{fig:dmanalysis}
\end{figure}

To elucidate the influence of Yukawa couplings and mass splitting on the relic abundance of dark matter, in Fig.~\ref{fig:dmanalysis},  we illustrate the variation of relic density with the dark matter mass. In the left panel of Fig.~\ref{fig:dmanalysis}, the Yukawa coupling $\kappa$ is varied within the range $\kappa \in [10^{-4},10^{-3}]$, while the mass difference between the lightest neutral scalar $\eta_R$ and $f$ (i.e., $M_{\eta_R}-M_f$) is varied in three different ranges, as indicated in the figure's inset. Evidently, an increase in the mass difference leads to a corresponding increase in the relic density. This trend arises because the co-annihilation contribution to $\langle \sigma v \rangle_{\rm eff}$ gradually diminishes with an increase in $(M_{\eta_R}-M_f)$, thereby boosting the relic abundance of $f$.

Expanding the analysis, in the right panel of Fig.~\ref{fig:dmanalysis}, the mass difference $(M_{\eta_R}-M_f)$ is varied within a small range of $[50,60]$ GeV and the variation of relic density with dark matter mass is then showcased for three different ranges of Yukawa couplings, as outlined in the figure's inset. It is evident that an increase in Yukawa coupling leads to a decrease in relic density, attributed to the increase in $\langle \sigma v \rangle_{\rm eff}$. Additionally, an intriguing observation from this figure is that, when $(M_{\eta_R}-M_f)\in [50,60]$ GeV and Yukawa couplings are small (i.e., $\kappa \lesssim \mathcal{O}(10^{-3})$), the relic density does not change with further reduction in Yukawa couplings, for dark matter masses exceeding 500 GeV, as indicated by the red and blue-colored points. This phenomenon can be explained by the fact that, in scenarios with small Yukawa couplings, neither the annihilation of DM nor the co-annihilation of DM with dark sector scalars efficiently affects the relic density. Instead, it is primarily determined by the co-annihilation contribution from the annihilation of dark scalar partners. Conversely, in situations where Yukawa couplings are large and the mass difference is substantial, the relic density is primarily influenced by the annihilation of DM, as indicated by the purple-colored points.

\begin{figure}[h]
    \centering
\includegraphics[scale=0.47]{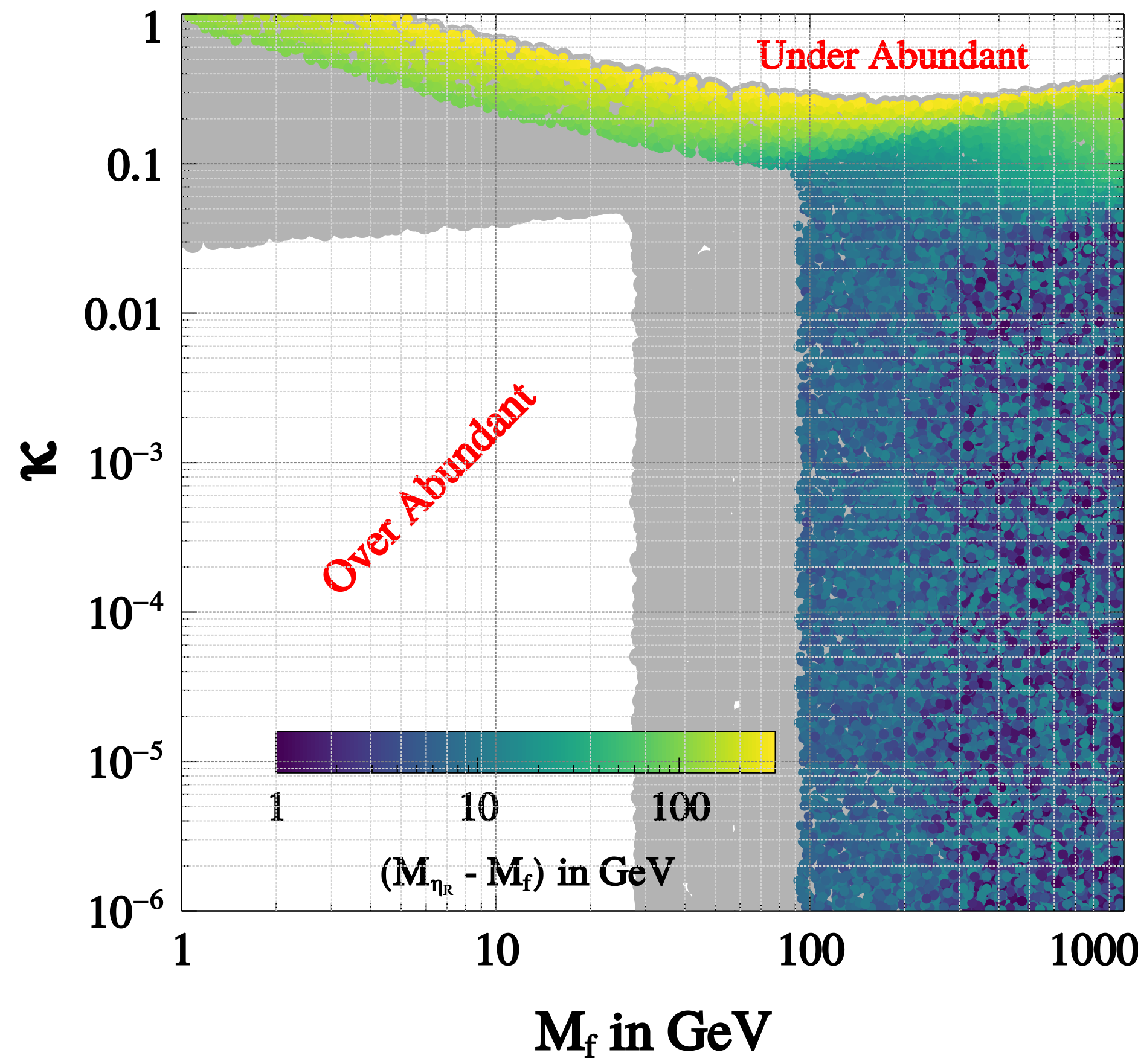}
    \caption{{Correct relic density satisfying points in the plane of DM mass $M_f$ and $\kappa$. The color code shows the value of  $M_{\eta_R}-M_f$}.}
    \label{fig:relicps}
\end{figure}

Thus, in summary of the effects that affect the relic density, in scenarios characterized by small Yukawa couplings and substantial mass differences ($\Delta_i$), the relic density is predominantly governed by the  co-annihilation contribution from the dark scalars. In such scenarios, DM annihilation becomes subdominant, and co-annihilation among dark matter and dark scalars is suppressed due to the large mass splitting. Conversely, when the mass difference between dark matter and dark scalars is not considerably large, co-annihilation among DM and dark scalars, as well as dark scalar annihilations, play a crucial role in determining the relic density. Only in cases where Yukawa couplings are significantly large, and the mass difference is also substantial, dark matter annihilations become relevant for achieving the correct relic density.

We present the parameter space satisfying correct relic density in the plane of $M_f$ and $\kappa$ with the color code representing the corresponding value of $M_{\eta_R}-M_f$ in Fig.~\ref{fig:relicps}. The grey-colored points are ruled out by imposing a conservative limit on the doublet scalar mass given by the LEP experiment of about $M_\eta  \geq 100$ GeV. It is evident that, when DM mass is small and $\kappa$ is small, the effective annihilation cross-section is very small and thus it is not possible to achieve correct relic density even with co-annihilation contributions. Thus we obtain a over abundant region below $M_f$ around $30$ GeV and $\kappa \lesssim \mathcal{O}(10^{-2})$. In the small DM mass range $M_f \lesssim 100$ GeV, correct relic density can be obtained only when the Yukawa couplings are significant {\it i.e.} $\kappa \sim \mathcal{O}(1)$ such that DM annihilation cross-section is appropriate to match the thermal cross-section as in this region the co-annihilation contributions are negligible. When DM mass is greater than $100$ GeV, and Yukawa coupling $\kappa \lesssim \mathcal{O}(10^{-2})$, we see that with increase in DM mass, the $M_{\eta_R}-M_f$ shows a gradual decrease to achieve the correct relic density. This is attributed to the fact that as the DM mass increases, the effective cross-section gradually decreases thereby increasing the relic density and thus it needs more effective co-annihilations which is possible by decreasing the $M_{\eta_R}-M_f$, to bring the relic density to correct ballpark. We also observe an under-abundant region when $\kappa \gtrsim \mathcal{O}(10^{-1})$ even with very large $M_{\eta_R}-M_f$. This is due to the fact that, with very large $\kappa$ the DM annihilation cross-section is large. So even if the co-annihilation contribution is suppressed because of large $M_{\eta_R}-M_f$, it is still not possible to achieve correct density.

\noindent
\underline{Direct Detection of DM:}\\ 
As the sole interaction connecting $f$ with SM particles is the Yukawa term in Eq.~(\ref{eq:Lag}), direct interactions between quarks and dark matter are absent at the tree-level. However, at the one-loop level, $f$ can have effective couplings with various SM particles, such as the photon, $Z$ boson, and Higgs boson.
Specifically, the exchange of the $Z$-boson results in the emergence of an effective axial-vector interaction which gives rise to a Spin-dependent DM-nucleon scattering and is dominant only when the couplings between Higgs and $\eta$ are very small. The constraints of spin dependent DM nucleon scattering is also relatively less constrained as compared to the spin-independent scattering cross-sections. Thus we focus here on the spin-independent DM-nucleon scattering rate as the direct-search experiments very stringently constrain it. 
\begin{figure}[h]
    \centering
    \includegraphics[scale=0.2]{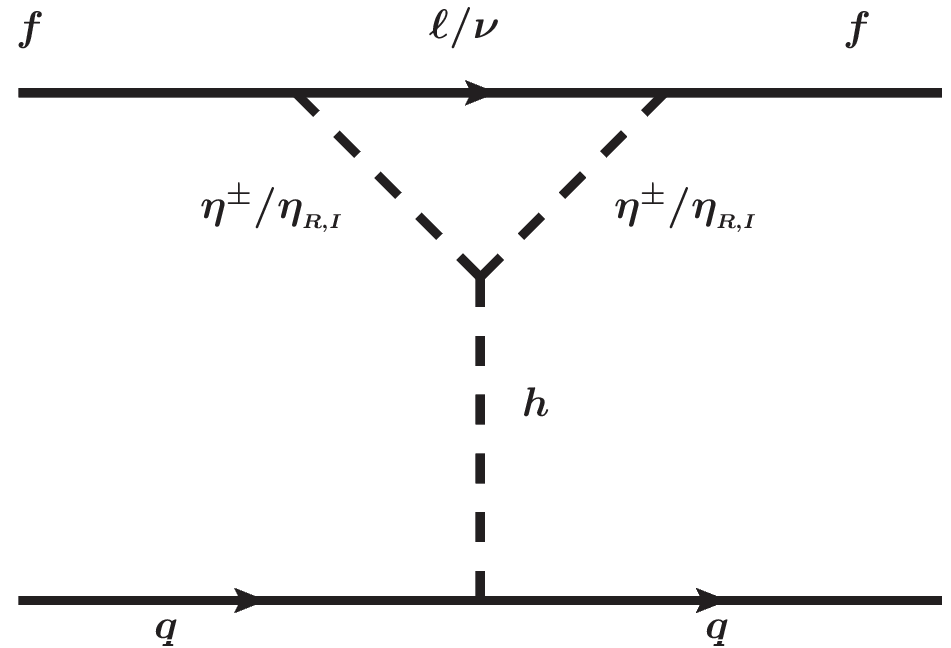}
    \caption{Spin-independent elastic DM-nucleon scattering arising at one loop.}
    \label{fig:ddprocess}
\end{figure}
The detection rate of dark matter particles within a detector can experience an amplification if the quartic couplings $\lambda_3$ and $\lambda_4$ are significant. When this condition is met, the exchange of Higgs bosons, as depicted in Fig.~\ref{fig:ddprocess}, leads to the emergence of an effective scalar interaction term between the quark $q$ and the dark matter particle $f$. This interaction is effectively described by $S_{q}~\bar{q}q~\bar{f} f$, where
\begin{align}
\label{eq:dd}
S_q &= -\frac{\kappa^2}{16\pi^2M_h^2M_f}\left[\lambda_3\mathcal{G}\left(\frac{M_f^2}{M_{\eta^\pm}^2}\right)+\frac{(\lambda_3+\lambda_4)}{2}\mathcal{G}\left(\frac{M_f^2}{M_{0}^2}\right)\right],
\end{align}
 with the loop function $\mathcal{G}(x)$ defined as: 
\begin{align}
\label{eq:G1_rephrased}
\mathcal{G}_1(x) &= \frac{x+(1-x)\ln(1-x)}{x},
\end{align}
and its value spans between 0 and 1 for $0\leq x\leq 1$.  
\begin{figure}[h]
    \centering
    \includegraphics[scale=0.45]{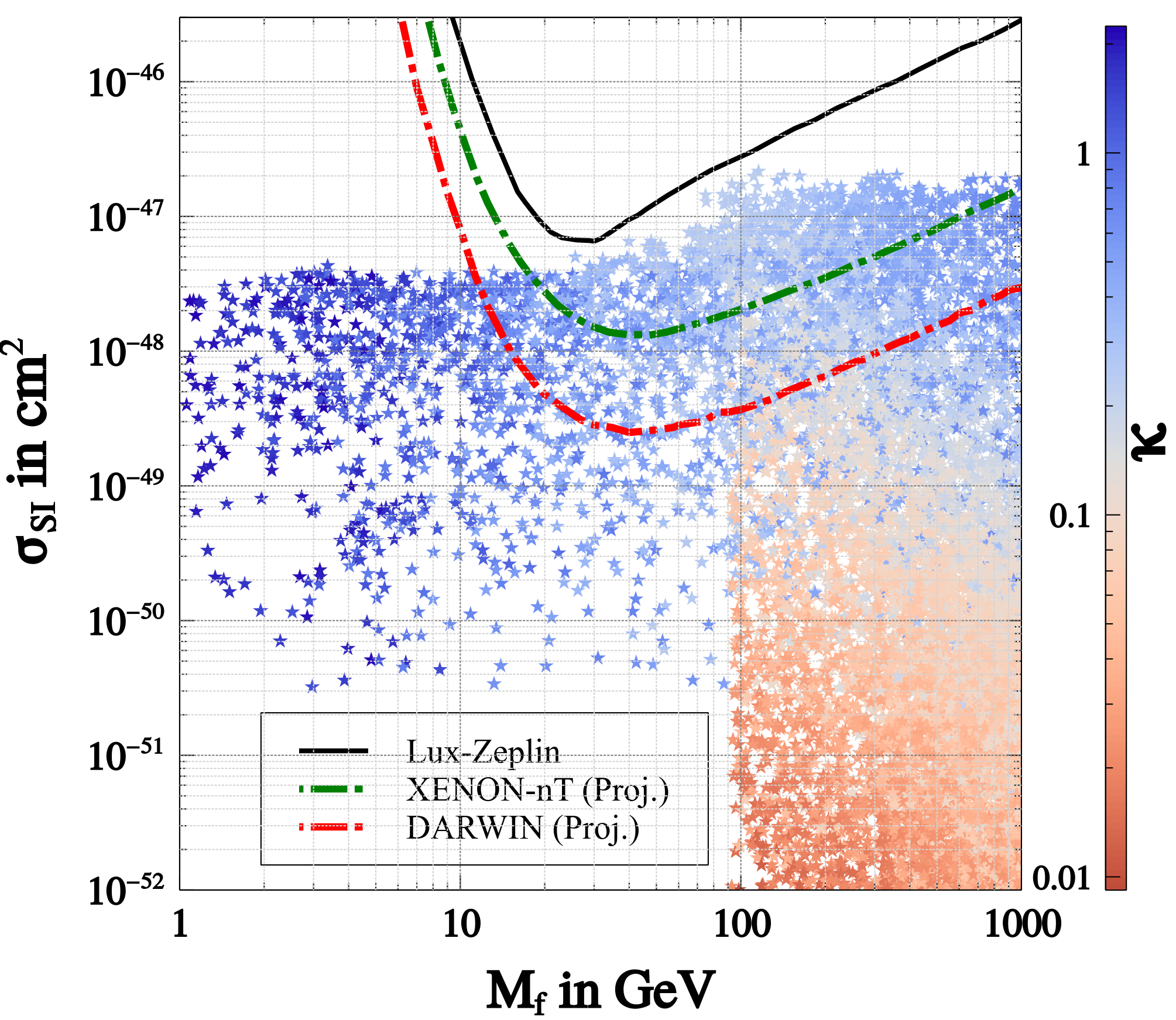}
    \caption{Spin-independent DM-nucleon scattering cross-section as a function of DM mass for the points satisfying correct relic density.  The color code represents the value of $\kappa$. The black solid line represents the most recent constraint from the Lux-Zeplin experiment. The green and red dot-dashed line represents the projected sensitivities of XENON-nT and DARWIN experiments, respectively.}
    \label{fig:dmdir}
\end{figure}
This interaction then results in the computation of the spin-independent cross section $\sigma_{SI}$ for the interaction of $f$ with a proton and expression for $\sigma_{\rm SI}$ is given by:
\begin{align}
\sigma_{\rm SI} &= \frac{4}{\pi}\frac{M_f^2m_p^2}{(M_f+m_p)^2}m_p^2 S^2_q f_p^2,
\end{align}
where $f_p$ represents the scalar form factor. We show the DM-nucleon scattering cross-section as a function of DM mass for the points satisfying correct relic density in Fig.~\ref{fig:dmdir}. Because of the loop suppression, we observe that even when the Yukawas and scalar quartic couplings are large, none of the points are ruled out, and the parameter space remains safe from the DM direct search constraints. However, interestingly, future experiments like XENON-nT~\cite{XENON:2020kmp} and DARWIN~\cite{DARWIN:2016hyl} with enhanced sensitivity can probe the Yukawa coupling $\kappa$ down to $\mathcal{O}(0.1)$.


\section{Higgs boson in the diphoton decay channel }\label{sec:Collider}
 The SM Higgs boson has a mass of $m_h \simeq 125$ GeV ~\cite{CMS:2021kom, ATLAS:2022tnm}, and one of the main decay channels  is the diphoton, where the SM rate for $h\to \gamma\gamma$ is dominated by the W-boson loop contribution.
 \begin{figure}[h]
	\begin{center}
		\includegraphics[scale=0.45]{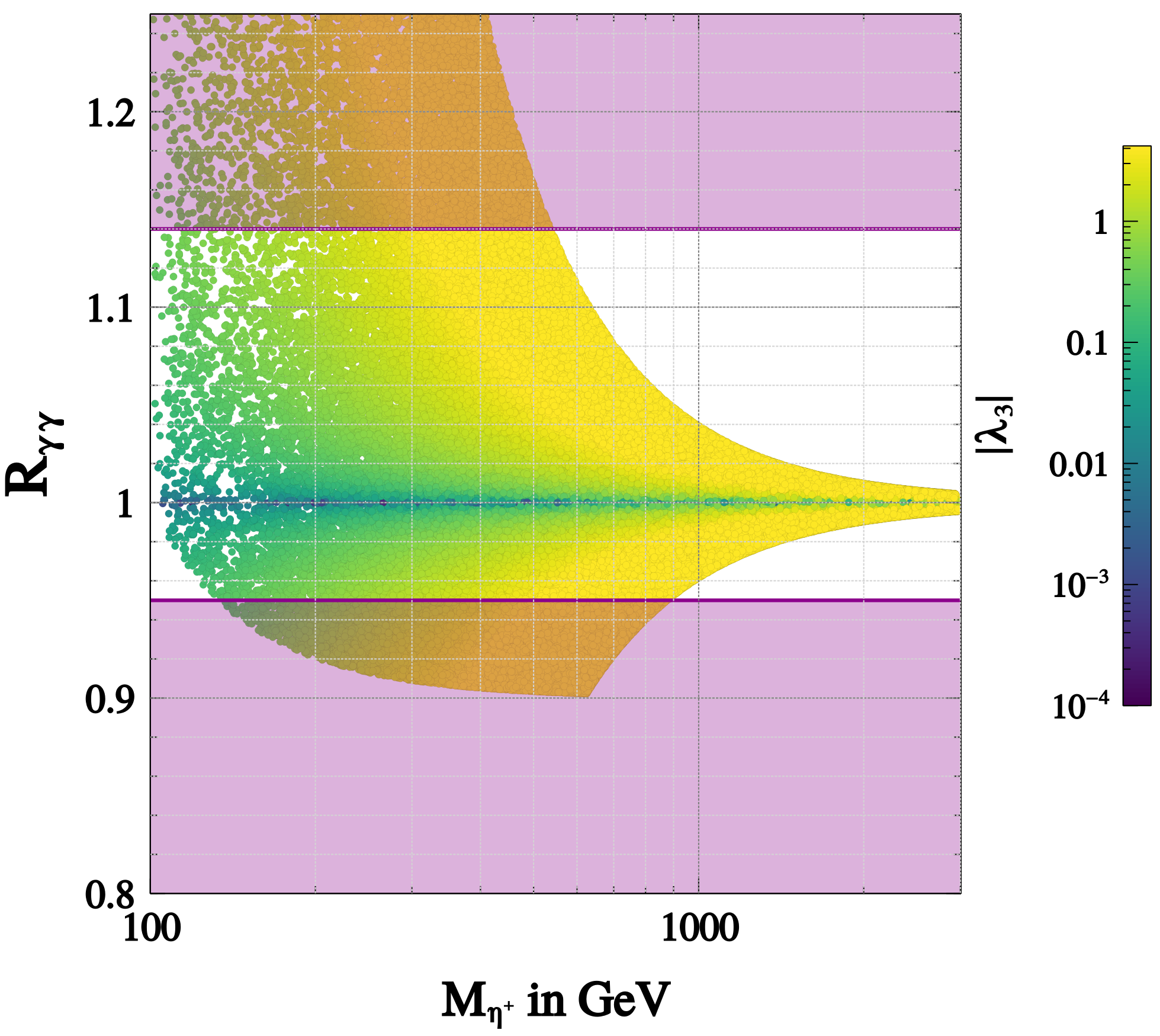}
	\end{center}
\caption{ $R_{\gamma\gamma}$ is plotted against $m_{\eta^+}$  in the plane of $\lambda_3$. The white-shaded region is the allowed region determined by the ATLAS experiment~\cite{ATLAS:2022tnm}}. 
\label{fig:h-signal-strength}
\end{figure}
The signal strength of $h\to \gamma\gamma$ is the ratio between the observed cross section $pp \to h \to \gamma\gamma $ and the same quantity computed in the SM. The observed cross section $pp\to h \to \gamma\gamma$ should match the FSS$_1$ model prediction. Since the dominant process of the Higgs boson production is the gluon fusion, in the first approximation, {the production cross-section of the Higgs boson in the FSS$_1$ model is the same as in the SM}. As a result, following \cite{Arhrib:2012ia}, after using narrow width approximation, the signal strength of $h \to \gamma\gamma$ in the model can be written as
 \begin{eqnarray}\label{eq:rgg}
    R_{\gamma\gamma}=\frac{\big[\sigma(gg\to h)\times{\rm Br}(h\to \gamma\gamma)\big]_{\rm FSS_1}}{\big[\sigma(gg\to h)\times{\rm Br}(h\to \gamma\gamma)\big]_{\rm SM}}
    =\frac{\Gamma_{\rm SM}^{h}}{\Gamma_{\rm FSS_1}^{h}}\frac{ \Gamma(h\to \gamma\gamma)_{\rm FSS_1}}{ \Gamma(h\to \gamma\gamma)_{\rm SM}}.
\end{eqnarray}
Here, the quantities with the FSS$_1$ and SM suffixes are computed in the flavored scoto-seesaw and the Standard Model, respectively. $\Gamma^{h}_{{{\rm FSS}_1},{\rm SM}}$ is the total decay width for these models. The  $h\to \gamma\gamma$ decay is experimentally well established, and in the LHC, the signal strength of $h\to \gamma\gamma$ is $R=1.04_{-0.09}^{+0.10}$ \cite{ATLAS:2022tnm}. While computing $R_{\gamma\gamma}$, we take for the total decay width of the Higgs boson $\Gamma_{\rm SM}^{h} = 4.07\times 10^{-3}$ GeV with a relative uncertainty of $^{+4.0\%}_{-3.9\%}$~\cite{ParticleDataGroup:2022pth}. For a theoretical error estimate, see also \cite{Freitas:2019bre}.
{For detailed studies on $h \to \gamma \gamma$  decays within  miscellaneous
beyond SM scenarios, see \cite{Cacciapaglia:2009ky, Posch:2010hx, Arhrib:2012ia, Hundi:2013lca, Hundi:2023tdq}}. In the framework of the FSS$_1$ model, this decay can be enhanced with the charged scalars ($\eta^{\pm}$) in the loop, over the SM contribution with charged fermions and $W$ bosons in the loop. Using Eq.~(\ref{eq:rgg}),  the expression for the partial decay width of $h\to \gamma\gamma$ in the FSS$_1$ model induced by the $\eta^{\pm}$ loop can be written as~\cite{Shifman:1979eb}
\begin{eqnarray}\label{eq:h-gg-model}
    \Gamma(h\rightarrow \gamma\gamma)=\frac{G_{F}\alpha_{em}^3 m_h^3}{128\sqrt{2}\pi^3}\Bigg| \sum_{f}N_f Q_f^2 F_{1/2}(\beta_f)+ F_1(\beta_W)+\frac{\lambda_3 v^2}{2M_{\eta^+}^2} F_0(\beta_{\eta^{\pm}})\Bigg|^2,
\end{eqnarray}
where $\beta_i=4 M_i^2/m_h^2$, $i=f,W,\eta^+$. $N_f$ is the color factor, and $Q_f$ is the charge of quarks. $\alpha_{em}$ and $G_{F}$ are the fine structure constant and Fermi constant. The $F$-functions in Eq. (\ref{eq:h-gg-model}) are the form factors of spin~$1/2, 1,0$ fields for the $h\to \gamma\gamma$ decay 
\begin{eqnarray}
		F_{1/2}(\beta_f)&=&-2\beta[1+(1-\beta)f(\beta)],\\
			F_1(\beta_W)&=& [2+ 3\beta+3 \beta(2-\beta)f(\beta)],\\
			F_0(\beta_{\eta^{\pm}})&=& \beta [1-\beta f(\beta)],
		\end{eqnarray}
  where
  \begin{eqnarray}
      f(\beta)&=&\big(\sin^{-1}\frac{1}{\sqrt{\beta}}\big)^2,\quad \beta \geq 1 \\
      &=&-\frac{1}{4} \Big[{\rm ln}\frac{1+\sqrt{1-\beta}}{1-\sqrt{1-\beta}}-i\pi \Big]^2, \quad \beta <1.
  \end{eqnarray}
 In the FSS$_1$ model, therefore,  the total decay width of the Higgs boson can be written as
\begin{eqnarray}\label{eq:gamma-FSS}
   \Gamma^{\rm h}_{{\rm FSS}_1}=\Gamma^{h}_{\rm SM}+\Gamma (h \to \eta_R \eta_R)+  \Gamma (h \to \eta_I \eta_I) + \Gamma (h \to \eta^+ \eta^-).
\end{eqnarray}
In the above equation, the decay width of the Higgs boson to different scalar particles is calculated using tree-level couplings 
\begin{eqnarray}
  &&  \lambda_{h\eta_R \eta_R}=\frac{2}{v}( M_{\eta_R}^2- \mu_2^2), \\ 
  && \lambda_{h\eta_I \eta_I}=\frac{2}{v}(M_{\eta_I}^2- \mu_2^2) .
\end{eqnarray}
Doing numerical analysis for $\Gamma(h\to \gamma\gamma)$, we scanned the parameters of the FSS$_1$ model in the range
\begin{eqnarray}
    100\hspace{1mm} \text{GeV} < M_{\eta_R},M_{\eta_I}, M_{\eta^+} < 2000 \hspace{1mm}\text{GeV},\quad |\lambda_{3,4,5}|\leq 4\pi.
\end{eqnarray}
In Eq.~(\ref{eq:gamma-FSS}), the total decay width of SM Higgs $h$ in the FSS$_1$ model has three extra contributions over the SM. In the FSS$_1$ framework, the scalars $\eta_R$ and $\eta_I$ are not the lightest neutral $Z_2$ odd particles of the theory, $f$ being the DM candidate. Thus with a judicious choice of the DM mass $M_f$ (satisfying relic density and direct search constraints), the Higgs boson decays to $\eta_R\eta_R$ and $\eta_I\eta_I$ can be made kinematically forbidden. The result of numerical analysis is shown in Fig.~\ref{fig:h-signal-strength} where the signal strength of $R_{\gamma\gamma}$ in the wide $\lambda_3$ range is given as a function of the charged scalar mass $M_{\eta^+}$.
The horizontal white region ($R_{\gamma\gamma}=1.04^{+0.10}_{-0.09}$) represents the currently allowed region measured by the ATLAS experiment using 139 fb$^{-1}$ of $pp$ collision data at $\sqrt s=13$ TeV~\cite{ATLAS:2022tnm}. This shows that the $M_{\eta^+}$ masses heavier than 1000 GeV are completely safe from LHC constraints. 
As follows from Eq.~(\ref{eq:h-gg-model}), if $\lambda_3<0$, the partial decay width of $h$ is smaller than in the SM while positive $\lambda_3$ will give the enhancement beyond the SM value. So, depending on the positive or negative values of $\lambda_3$, we get $R_{\gamma\gamma}>1$ or $R_{\gamma\gamma}<1$, respectively.  This behaviour can be seen in Fig.~\ref{fig:h-signal-strength}.

\section{Lepton Flavor Violation}\label{sec:LFVpheno}
The constraints on lepton flavor violating processes is an important aspect of the FSS$_1$ model under consideration. The model offers specific predictions, given that the flavor structure of the Yukawa couplings is entirely dictated by the $A_4$ discrete flavor symmetry and the alignment of flavon vacuums. Along with neutrino masses, mixing and DM phenomenology, LFV decays also give valuable insight on the FSS$_1$ model parameters. As a consequence of the considered flavor symmetry, the Yukawa couplings in the charged lepton sector are diagonal, see Eq. (\ref{eq:cl-mass-matrix}). However, Yukawa couplings $y_N$ and $y_s$ in Eq.~(\ref{eq:Lag}) associated with the type-I seesaw and scotogenic mechanisms, respectively, contribute to the LFV decays. These Yukawa couplings can generate lepton flavor violating processes like $l_{\alpha}\to l_{\beta}\gamma$ and $l_{\alpha}\to 3 l_{\beta}$ ($\alpha,\beta = e, \mu,\tau$)\footnote{ To study lepton flavor violation in the pure scotogenic model, see Refs.~\cite{Toma:2013zsa,Vicente:2014wga,Hagedorn:2018spx}.}. Studies on these LFV decays completely depend on the FSS$_1$ model construction as described below.

In our framework, the branching ratios of the  $l_{\alpha}\rightarrow l_{\beta}\gamma$ decays for the scotogenic contribution can be written as~\cite{Rojas:2018wym,Toma:2013zsa}
\begin{eqnarray}
    {\rm Br}(l_{\alpha} \rightarrow l_{\beta}\gamma) \approx \frac{3 \pi{\alpha_{em}}}{64 G_F^2}|Y_F^{\beta *}Y_{F}^{\alpha}|^2 \frac{1}{M_{\eta^+}^4}\Bigg(F_1\Bigg(\frac{M_f^2}{M_{\eta^+}^2}\Bigg)\Bigg)^2 {\rm Br}(l_{\alpha} \rightarrow l_{\beta}\nu_{\alpha}\bar{\nu}_{\beta}).
\end{eqnarray}
Here $G_F$ is the Fermi constant,  $Y_F$ is the Yukawa coupling matrix from the scotogenic contribution given in Eq. (\ref{eq:scoto yukawa}). The expression for the function $F_1$ is given by
\begin{eqnarray}
    F_1(x)=\frac{1-6x+3x^2+2 x^3-6 x^2 {\rm log}x}{6(1-x)^4}.
\end{eqnarray}
As mentioned earlier, the Yukawa couplings are determined by the considered discrete symmetries of the model. Due to the specific VEV structure of the $A_4$ triplet flavon, $Y_F^{\tau}=0$ as given in Eq.~(\ref{eq:scoto yukawa}). Therefore, the scotogenic contribution alone yields a vanishing contribution for $\tau \rightarrow e\gamma$ and $\tau \rightarrow \mu\gamma$ lepton flavor violating decays. So, the only non-vanishing contribution arising in the decays of the form $\l_{\alpha}\rightarrow \l_{\beta}\gamma$ comes from the $\mu \rightarrow e\gamma$ decay, the expression of branching fraction of which is given by~\cite{Rojas:2018wym,Toma:2013zsa}
\begin{eqnarray}\label{eq:scoto-mu-e-gamma0}
    {\rm Br}(\mu \rightarrow e \gamma) &\approx & \frac{3 \pi{\alpha_{em}}}{64 G_F^2}|2y_s y_s^* \epsilon^4|^2 \frac{1}{M_{\eta^+}^4}\Bigg(F_1\Bigg(\frac{M_f^2}{M_{\eta^+}^2}\Bigg)\Bigg)^2 {\rm Br}(\mu \rightarrow e\nu_{\mu}\bar{\nu}_{e})\\ 
   & = &\frac{3 \pi {\alpha_{em}}}{16 G_F^2M_f^2}\Bigg(\frac{|b|}{\mathcal{F}(M_{\eta_R},M_{\eta_I},M_f)}\Bigg)^2 \frac{1}{M_{\eta^+}^4}\Bigg(F_1\Bigg(\frac{M_f^2}{M_{\eta^+}^2}\Bigg)\Bigg)^2 {\rm Br}(\mu \rightarrow e\nu_{\mu}\bar{\nu}_{e}),\label{eq:scoto-mu-e-gamma1}
\end{eqnarray}
where we have substituted Eq.~(\ref{eq:scoto yukawa}) in Eq.~(\ref{eq:scoto-mu-e-gamma0}) to obtain Eq.~(\ref{eq:scoto-mu-e-gamma1}). In the above, $\epsilon=v_f/\Lambda$, where for simplicity we have assumed all flavon VEVs to be the same, $i.e.$ $v_{\xi}=v_{s,a}=v_f$. Clearly, ${\rm Br}(\mu \to e\gamma)$ depends on the parameter $|b|$, which is constrained from neutrino oscillation data ranging from 0.0048 to 0.0056 eV as given in Fig. \ref{fig:alpha}. 
\begin{figure}
    \centering
    \includegraphics[scale=0.43]{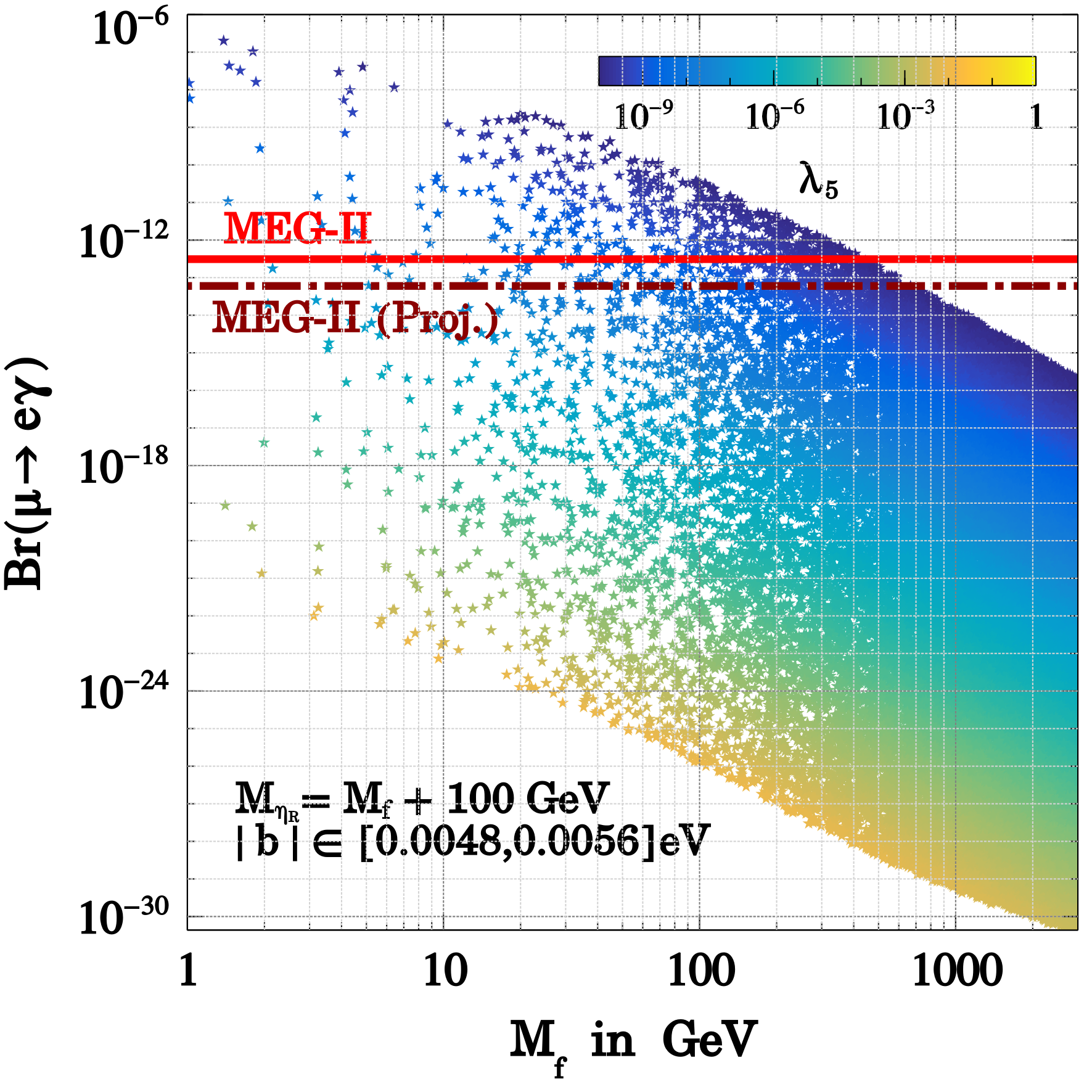}
    \includegraphics[scale=0.43]{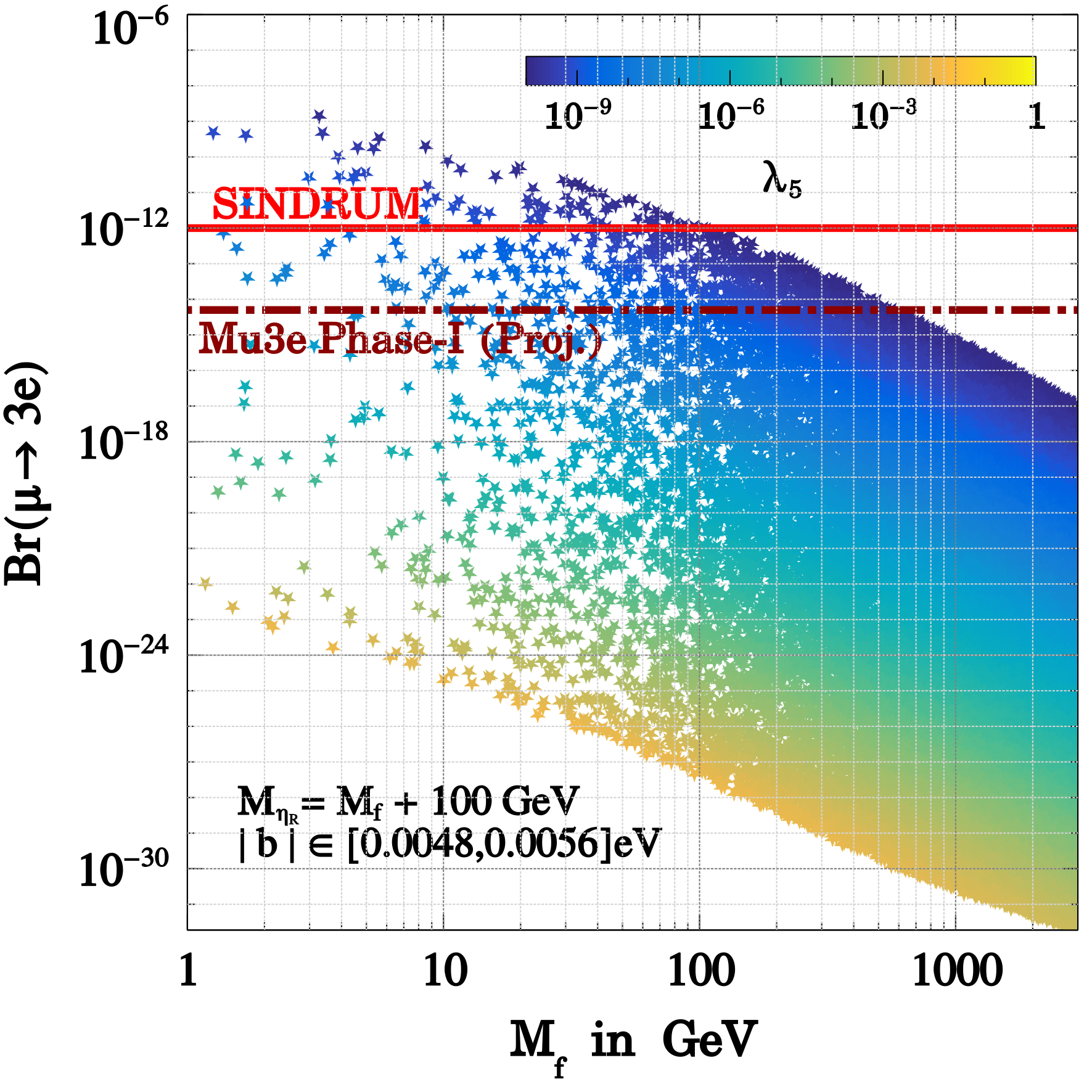}
    \caption{Branching ratio of $\mu \to e\gamma$ ($\mu \to 3e$) is plotted against the dark matter mass $M_f$ in the left (right) panel. In both panels, the red line corresponds to the present upper bound and the pink line corresponds to the future sensitivity (see text for details). }
    \label{fig:mu-eg-mu-3e}
\end{figure}
\begin{figure}[h]
   \centering
   \includegraphics[scale=0.5]{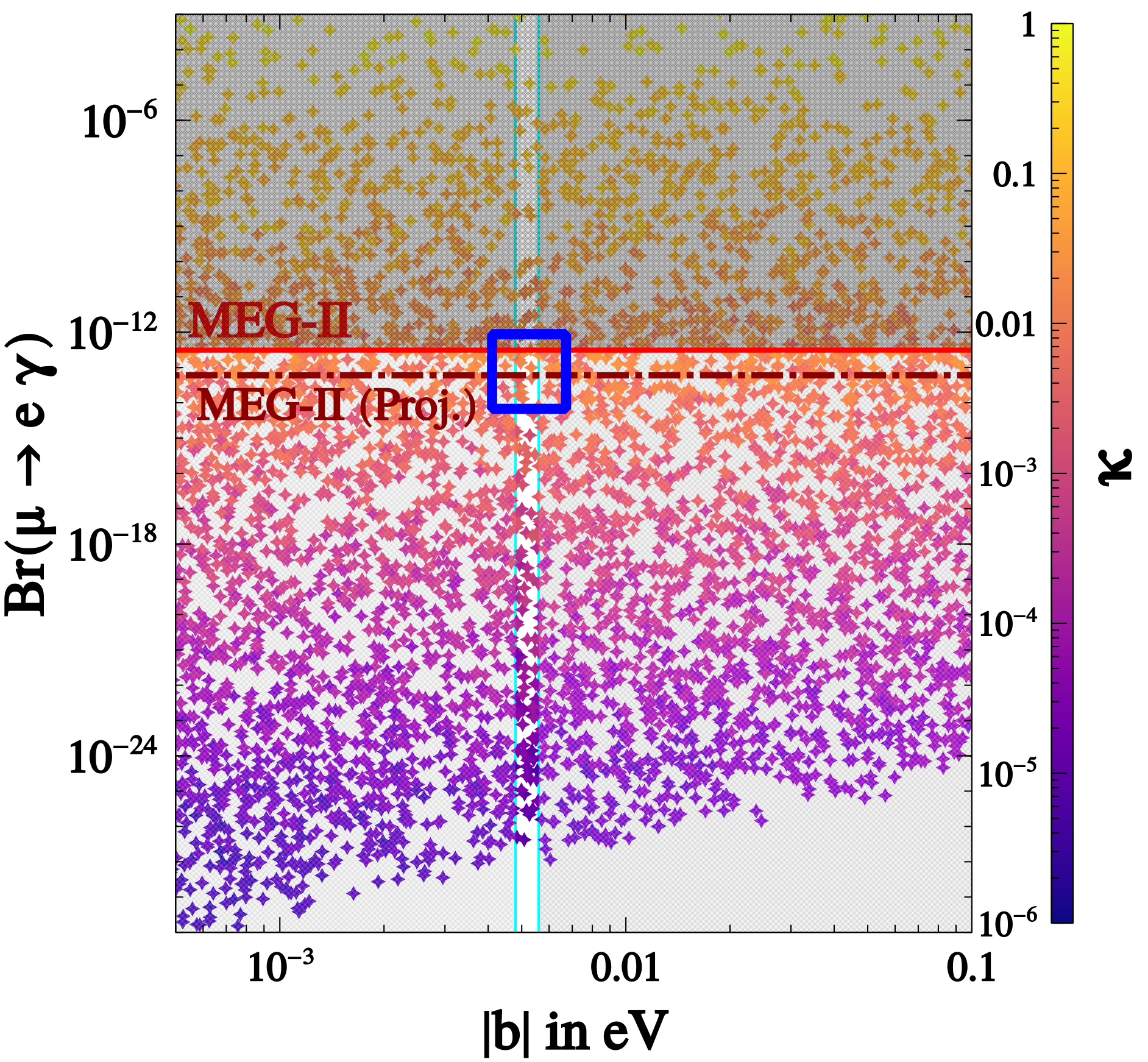}
   \caption{Branching ratio for $\mu \to e \gamma$ is plotted against $|b|$ for the points satisfying correct relic density of DM. The variation of color legends is for the variation of $\kappa$. Cyan shaded regions rules-out values of $|b|$ other than that consistent with the neutrino oscillation data $(0.0048-0.0056)$ eV. Solid red and dot-dashed brown lines represent the recent updated constraint from MEG-II and its projected sensitivity, respectively.}
   \label{fig:br-mu-eg-b}
\end{figure}
Another type of the LFV decays appearing in our FSS$_1$ framework are the $l_{\alpha}\rightarrow 3 l_{\beta}$ ($\l_{\alpha} \rightarrow l_{\beta}\bar{l}_{\beta}l_{\beta}$) processes. The corresponding branching ratios are given by~\cite{Toma:2013zsa}
\begin{eqnarray}
    {\rm Br}(l_{\alpha}\rightarrow 3 l_{\beta})\approx \frac{3{\alpha_{em}}^2}{512 G_F^2}|Y_F^{\beta ^*}Y_F^{\alpha}|^2 \frac{1}{M_{\eta^+}^4}\mathcal{G}\Big(\frac{m_{\alpha}}{m_{\beta}}\Big)\Bigg(F_2\Bigg(\frac{M_f^2}{M_{\eta^+}^2}\Bigg)\Bigg)^2 {\rm Br}(l_{\alpha} \rightarrow l_{\beta}\nu_{\alpha}\bar{\nu}_{\beta}),
\end{eqnarray}
where
\begin{eqnarray}
    F_2(x)&=&\frac{2-9x+18x^2-11x^3+6x^3 {\rm log}x}{6(1-x)^4},
\\
    \mathcal{G}\Big(\frac{m_{\alpha}}{m_{\beta}}\Big)&=&\Big(\frac{16}{3}{\rm log}\Big(\frac{m_{\alpha}}{m_{\beta}}\Big)-\frac{22}{3}\Big).
\end{eqnarray}
Again, following Eq.~(\ref{eq:scoto yukawa}),  we find that $Y_F^{\tau}=0$,  hence the branching fractions for $\tau \rightarrow 3e$ and $\tau \rightarrow 3\mu$ decays vanish. The only non-vanishing contribution originates from the $\mu \rightarrow 3 e$ decay, and the branching fraction can be written as~\cite{Toma:2013zsa}
\begin{eqnarray}\label{eq:scoto-mu-3e0}
  {\rm Br}(\mu\rightarrow 3 e)&\approx &\frac{3{\alpha_{em}}^2}{512 G_F^2}|2 y_s y_s^* \epsilon^4|^2 \frac{1}{M_{\eta^+}^4}\mathcal{G}\Big(\frac{m_{\mu}}{m_{e}}\Big)\Bigg(F_2\Bigg(\frac{M_f^2}{M_{\eta^+}^2}\Bigg)\Bigg)^2, \\
 &=&\frac{3\alpha_{em}^2}{128 M_f^2 G_F^2} \Bigg(\frac{|b|}{\mathcal{F}(M_{\eta_R},M_{\eta_I},M_f)}\Bigg)^2 \frac{1}{M_{\eta^+}^4}\mathcal{G}\Big(\frac{m_{\mu}}{m_{e}}\Big)\Bigg(F_2\Bigg(\frac{M_f^2}{M_{\eta^+}^2}\Bigg)\Bigg)^2,\label{eq:scoto-mu-3e1}
\end{eqnarray}
where we have substituted Eq.~(\ref{eq:scoto yukawa}) in Eq.~(\ref{eq:scoto-mu-3e0}) to obtain Eq.~(\ref{eq:scoto-mu-3e1}) with $\epsilon=v_f/\Lambda$. Similar to Eq.~(\ref{eq:scoto-mu-e-gamma1}), here we also find that ${\rm Br}(\mu\rightarrow 3 e)$ depends on the scotogenic mass parameters $M_{f,{\eta^+},{\eta_R},{\eta_{I}}}$ as well as $|b|$, the parameter involved in explaining correct neutrino oscillation parameter and DM relic density. The variation of the corresponding coupling $\lambda_5$ is given in the inset. In Fig.~\ref{fig:mu-eg-mu-3e}, we have shown plots for  $\mu \to e\gamma$ (left panel) and $\mu \to 3 e$ (right panel) branching ratios against the dark matter mass $M_f$, satisfying bound on $|b|$ obtained from Fig.~\ref{fig:alpha}.  The current constraints (denoted by magenta lines) for the branching ratio of the  $\mu \to e\gamma$ decay is given by the MEG-II experiment as ${\rm Br}(\mu \to e\gamma) \leq 3.1 \times 10^{-13}$ \cite{MEGII:2023ltw} whereas for $\mu\rightarrow 3 e$ decay the   constraint by SINDRUM experiment is given as ${\rm Br}(\mu\rightarrow 3 e) \leq 1 \times 10^{-12} $~\cite{SINDRUM:1987nra}. In both plots, a current upper bound on both these decays, constrains the dark matter mass $M_f$ specifically in the low mass region. $M_f$ will be further constrained by the future MEG-II (Proj.)~\cite{MEGII:2018kmf} and Mu3e Phase-I ~\cite{Perrevoort:2018ttp} experiments.  To illustrate the dependence of the LFV branching ratio on the neutrino oscillation parameters and its consistency with the DM phenomenology, in Fig.~\ref{fig:br-mu-eg-b}, we have plotted  ${\rm Br}(\mu \rightarrow e \gamma)$ against $|b|$. Here, the white-shaded region is consistent with correct neutrino masses and mixing given in the right panel of Fig.~\ref{fig:alpha}.  Hence,  the cyan-shaded regions are ruled out by neutrino oscillation data. This plot also depicts the dependence of the branching ratio on the scotogenic Yukawa coupling shown by the variation of $\kappa$. The upper shaded region is already ruled out by the recent updated constraint from MEG-II~\cite{MEGII:2023ltw}, the projected sensitivity of MEG-II can probe $\kappa$ of the order $\mathcal{O}(10^{-2})$.
\begin{table}[h!]
    \centering
    \begin{tabular}{|c|c|c|c|}
    \hline
      Decay Modes   & Scotogenic  & Seesaw  & Remarks \\
      \hline
       $\mu \to e\gamma$ &\checkmark &\ding{55} & $Y_N^e=0$ \\ 
       $\tau \to e\gamma$ &\ding{55} &\ding{55} & $Y_F^{\tau}=0,Y_N^e=0$\\
       $\tau \to \mu\gamma$  &\ding{55} & \checkmark & $Y_F^{\tau}=0$\\ 
       $\mu \to 3e$ &\checkmark &\ding{55}  & $Y_N^e=0$\\ 
       $\tau \to 3e$  &\ding{55} &\ding{55} & $Y_F^{\tau}=0,Y_N^e=0$\\
       $\tau \to 3\mu$  &\ding{55} & \checkmark & $Y_F^{\tau}=0$\\
       \hline
    \end{tabular}
    \caption{Possible lepton flavor violating decay modes and their origin in the FSS$_1$ framework.  
    The \checkmark  and \ding{55} symbols stand for allowed and disallowed regimes.}
    \label{tab:LFV}
\end{table}

For the type-I seesaw contributions to LFV decays,  the  branching fractions for $l_{\alpha}\to l_{\beta}\gamma$ decays can be cast in the following form  
\begin{eqnarray}\label{eq:lfv1 seesaw}
   {\rm Br}(l_{\alpha} \rightarrow l_{\beta}\gamma)&\approx&    \frac{3 {\alpha_{em}} v^4}{8 \pi M_{N}^4} \Bigg| Y_N^{\beta} Y_N^{\alpha^*} f\Bigg(\frac{M^2_{N}}{M_W^2}\Bigg)\Bigg|^2,
\end{eqnarray}
where $Y_N$ is given in Eq.~(\ref{eq:seesaw yukawa}). The loop function $f(x)$ in Eq.~(\ref{eq:lfv1 seesaw}) is
\begin{eqnarray}
    f(x)=\frac{x(2x^3+3x^2-6x-6 x^2{\rm log}(x) +1)}{2(1-x)^4}. 
\end{eqnarray}
Similar to the scotogenic contribution, the $A_4$ discrete symmetry and the VEV alignment of the flavon $\phi_s$ plays a crucial role in estimating the branching ratio for $l_{\alpha}\rightarrow l_{\beta}\gamma$. The VEV alignment of the flavon $\phi_s$ is such that it gives $Y_N^e=0$ as a result of which the branching fraction for $\mu \rightarrow e\gamma$ and $\tau \rightarrow e\gamma$ decays vanish. The only non-vanishing contribution arising in the type of $l_{\alpha}\rightarrow l_{\beta}\gamma$ decay is $\tau \rightarrow \mu \gamma$ and the branching fraction is given by
\begin{eqnarray}\label{eq:br-tau-seesaw}
   {\rm Br}(\tau \rightarrow \mu\gamma)&=&\frac{3 {\alpha_{em}} v^4}{8 \pi M_{N}^4} \Bigg| (y_N y_N^{*} \epsilon^2 f\Bigg(\frac{M^2_{N}}{M_W^2}\Bigg)\Bigg|^2 =\frac{3{\alpha_{em}}}{8\pi M_N^2}|a|^2\Bigg|f\bigg(\frac{M_N^2}{M_W^2}\bigg)\Bigg|^2.
\end{eqnarray}
For $M_N \sim 10^{4}$ GeV and $|a|=0.0250$ eV, the branching fraction in Eq.~(\ref{eq:br-tau-seesaw}) gives $5.4 \times 10^{-33}$ which is very small compared to the experimental  limit~( $4.4 \times 10^{-8}$)~\cite{BaBar:2009hkt}. For higher $M_N$ values, the branching ratio will be even more suppressed.  Similarly, the branching ratio for the $\tau\rightarrow 3 \mu$ conversion is found to be very small compared to the experimental bound~\cite{Hayasaka:2010np}.

In Tab.~\ref{tab:LFV}, we have summarized the allowed  LFV decays in the FSS$_1$ model. The considered discrete flavor symmetry and corresponding vacuum alignment of the flavons completely disallow decay channels such as $\tau\rightarrow e\gamma$ and $\tau \rightarrow 3e$. Such a decisive prediction can be made since we have vanishing values for the  Yukawa couplings  $Y_F^{\tau}$ and $Y_N^e$, see Eq.~(\ref{eq:seesaw yukawa}) and Eq.~(\ref{eq:scoto yukawa}), associated with the scotogenic and type-I seesaw contributions, respectively. Present experiments already exclude branching ratios larger than about $\mathcal{O} (10^{-8})$. Any positive signal by the future experiments will essentially test the validity of the FSS$_1$ framework. 
 

\section{Summary of Phenomenological Analysis}\label{sec:summary}
Both type-I seesaw and the scotogenic contribution within our FSS$_1$ framework are crucial in explaining the hierarchy associated with the neutrino mass-squared differences. The scotogenic contribution is characterized by the parameter $b$ and its magnitude is restricted within a narrow range   $0.0048-0.0056$ eV.
\begin{figure}[h]
    \centering
    \includegraphics[scale=0.5]{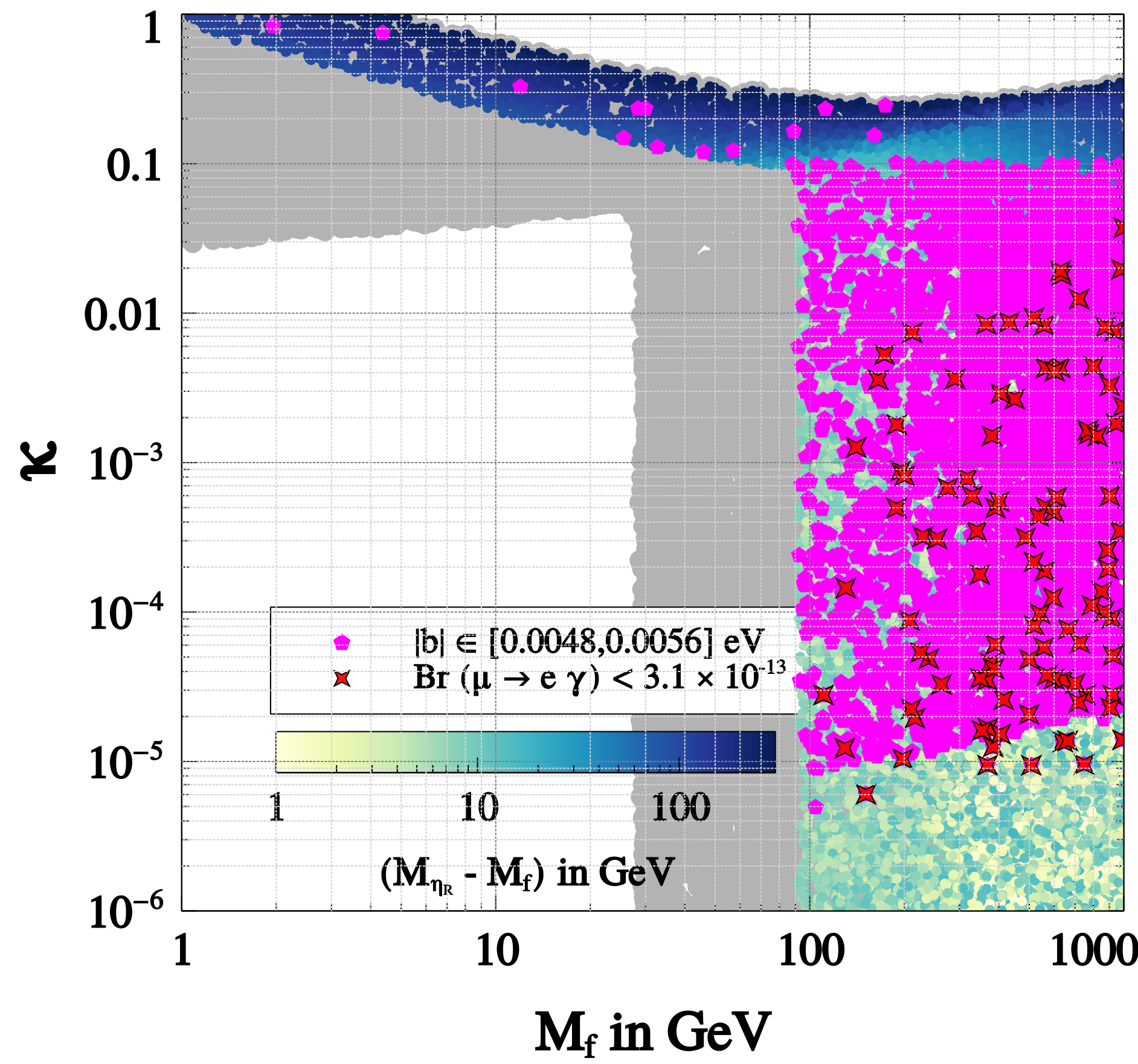}
    \caption{Final parameter space for the two important parameters of the FSS$_1$ model, namely the Yukawa coupling $\kappa$ and the DM fermion  mass $M_f$, after imposing the constraints from DM relic density, direct search, neutrino oscillation data, and LFV {decays}. }
    \label{fig:all}
\end{figure}
The estimation of  DM relic density depends on the scotogenic contribution Yukawa coupling $\kappa$ associated with $|b|$ as given in Eq. (\ref{eq:scoto yukawa}). This dependence is shown in Fig.~\ref{fig:relicps} in the DM mass $M_f - \kappa$ plane. The allowed parameter space gets further constrained to satisfy correct neutrino oscillation data and experimental limits on LFV decays discussed in Section \ref{sec:nu-mixing} and \ref{sec:LFVpheno},  respectively. Although the allowed range of $|b|$ is tightly constrained from neutrino oscillation data, interplay of DM $f$ and other dark sector particles $\eta_{I,R,\pm}$ can satisfy correct DM relic density with contributions from various annihilation and co-annihilation contributions mentioned in Fig. \ref{fig:dmannfeyn}~-~\ref{fig:scalarcoannfeyn}.  Hence, updating Fig. \ref{fig:relicps}, in Fig.~\ref{fig:all}, we have plotted the final parameter space, which includes constraints from DM relic density,  neutrino oscillation data, and LFV decays. The points with the color code represent the parameter space consistent with the DM relic density and direct search constraints. Once we impose the constraints for $|b|$ from the neutrino oscillation data obtained from Fig. \ref{fig:alpha}, we get the magenta-colored points. Finally, we obtain the red star points when we impose the constraint from LFV decays along with the constraints mentioned above from DM phenomenology and neutrino oscillation. LFV constraints restrict the maximum allowed Yukawa coupling to be less than $\mathcal{O}(10^{-2})$ and DM masses between $(100-1000)$ GeV are found to be simultaneously consistent with neutrino oscillation data,  DM relic density, direct search, and LFV decay constraints.

\section{Conclusions and Outlook}\label{sec:conc}

We propose the flavor-scoto-seesaw (FSS) model, which explains the observed hierarchy between the solar and atmospheric neutrino mass scales, experimentally allows the trimaximal mixing scheme, and naturally accommodates viable dark matter candidates. In this framework, type-I seesaw and 1-loop scotogenic mechanisms contribute to the effective light neutrino mass.  With only one right-handed neutrino, the type-I seesaw contribution dominantly contributes to generating atmospheric neutrino mass scale, and the scotogenic contribution (with the involvement of the dark fermion $f$ and scalar $\eta$) is mainly responsible for the solar neutrino mass scale. The whole framework is embedded within $A_4\times Z_4 \times Z_3 \times Z_2$ discrete flavor symmetry predicting the lightest neutrino to massless and one non-vanishing Majorana phase. The model also contains a few flavon fields to realize appropriate flavor structure to explain observed neutrino mixing.  The inclusion of auxiliary $Z_N$ ($N=$ 4, 3, 2) symmetries is a generic feature of discrete flavor symmetric models to forbid several unwanted terms, and the charged lepton mass matrix is found to be a diagonal one. These $Z_N$ symmetries are broken down to a dark $Z_2$ symmetry, ensuring the stability of dark matter under which only $f$ and $\eta$ are odd. With a judicious choice of the flavon vacuum alignments,  the TM$_1$ mixing scheme can be realized, and hence we call our flavor symmetric scoto-seesaw model an FSS$_1$ model. 

Considered flavor symmetry completely dictates the flavor structure of the model and makes it highly predictive. The FSS$_1$ model provides rich phenomenology for neutrino masses, mixing, LFV decays, and collider studies and accommodates potential dark matter candidates with DM $f$ fermion and $\eta$ scalar. With both type-I and scotogenic contributions, a rank-2 light neutrino mass matrix is obtained, predicting normal ordering of light neutrino mass. The presence of flavor symmetry in FSS$_1$ implies a preference for the higher octant of the atmospheric mixing angle $\theta_{23}$ where the allowed ranges given by $0.531\leq\sin^2\theta_{23}\leq 0.544$ and $0.580\leq\sin^2\theta_{23}\leq 0.595$. The model also tightly constrains the TM$_1$  prediction for the Dirac CP phase $\delta_{\rm CP}$ (within the range $\pm(1.44-1.12)$ radian) and the Jarlskog CP invariant.  Moreover, correlations among neutrino mixing parameters within the FSS$_1$ model (see Figs.~\ref{fig:ss23-dcp} and \ref{fig:majorana phase}) give a strict determination of the Majorana CP phase, thus giving an accurate prediction for $m_{\beta \beta}$ (see Table~\ref{tab:mass})  within the range $1.61-3.85$ meV. Here, the dark fermion $f$ is considered as the DM candidate whose production mechanism is connected with its Yukawa coupling with SM leptons and the inert doublet scalar $\eta$. The magnitude of these Yukawa couplings plays a critical role in determining correct neutrino mixing and DM relic density through the thermal freeze-out mechanism.
With the flavor structure of the FSS$_1$ framework, only the scotogenic part contributes to the lepton flavor violating decays such as $\mu \rightarrow e\gamma, \mu \rightarrow 3e$. On the other hand, though they are very small, the seesaw part of FSS$_1$ only contributes to decays such as $\tau \rightarrow \mu\gamma, \tau \rightarrow 3 \mu$.  Interestingly, owing to the flavor symmetry and vacuum alignment of the flavons, LFV decays such as $\tau\rightarrow e\gamma$ and $\tau\rightarrow 3 e$ are completely disallowed, and any positive signal LFV for these two decays will test the viability of this model. Within the FSS$_1$ framework, the WIMP  DM masses between $(100-1000)$ GeV are simultaneously consistent with the constraints from neutrino oscillation data,  DM relic density, direct search, and LFV decays.  

The FSS$_1$ model can also be tested at the colliders via a wide range of phenomenological studies. For example, FSS$_1$ can contribute to the Higgs boson diphoton decay channel $h \to \gamma \gamma $. Fig.~\ref{fig:h-signal-strength} shows that $M_{\eta^+}$ masses up to 1 TeV can have implications at the diphoton Higgs decay channel using the present LHC experimental results. With the increasing data collection at LHC and HL-LHC, the precision of $R_{\gamma\gamma}$ will improve, giving prospects for better determination of allowed regions for specific flavor model parameters. Thus, phenomenology-based $R_{\gamma\gamma}$ constraints can be used for further studies and predictions for producing exotic discrete flavor model signals at present and future colliders. The same statement is valid for further phenomenological studies of the model based on DM and LFV constraints. Thus, in alignment with all pertinent constraints, the model retains its predictiveness across LFV experiments, direct detection of DM, as well as collider experiments.

\section*{}
 \appendix
 
\section{Appendix: $A_4$ symmetry}\label{section:A4 group}
$A_4$ is a discrete group of even permutations of four
objects\footnote{For a detailed discussion on $A_4$ see Refs.~\cite{Altarelli:2010gt,Ishimori:2010au}}. Geometrically, it is an invariance group of a tetrahedron. It has 12 elements which can be generated by two basic objects $S$ and $T$ which obey the following relations
\begin{eqnarray}
	S^2=T^3=(ST)^3=1
\end{eqnarray}
The $A_4$ group has three one-dimensional irreducible representations $1$,$1^{\prime}$ and $1^{\prime\prime}$ and one three dimensional irreducible representation 3. Products of the singlets and triplets are
given by~\cite{Ishimori:2010au}
\begin{eqnarray}
	1&\otimes& 1=1; ~  
	1^{\prime}\otimes1^{\prime\prime}=1,  \\
	1^{\prime}&\otimes& 1^{\prime}=1^{\prime\prime};~  
	1^{\prime\prime}\otimes 1^{\prime\prime}=1^{\prime}, \\
	3&\otimes&3 =1\oplus1^{\prime}\oplus 1^{\prime\prime}\oplus 3_s\oplus 3_a,
\end{eqnarray}
where the subscripts $``s"$ and $``a"$ denote symmetric and antisymmetric parts, respectively. Writing two $A_4$ triplets  as $X=(x_1,x_2,x_3)^{T}$ and $Y=(y_1,y_2,y_3)^T$ respectively, their product can be written as~\cite{Ishimori:2010au} 
\begin{eqnarray}
        X \otimes Y&=&(X \otimes Y)_{1}\oplus(X \otimes Y)_{1'}\oplus(X \otimes Y)_{1''}\oplus(X \otimes Y)_{3s}\oplus(X \otimes Y)_{3a}
\end{eqnarray} 
where
\begin{eqnarray}        
	(X \otimes Y)_{1}&\sim &x_1 y_1+x_2y_3+x_3 y_2,\\
	(X \otimes Y)_{1'}&\sim &x_3 y_3+x_1y_2+x_2 y_1,  \\
	(X \otimes Y)_{1''}&\sim &x_2 y_2+x_1y_3+x_3 y_1,  
 \end{eqnarray} 
 
\begin{eqnarray} 
	(X \otimes Y)_{3s}&\sim& \begin{pmatrix}
		2 x_1 y_1-x_2 y_3-x_3y_2 \\
		2 x_3 y_3-x_1 y_2-x_2 y_1 \\
		2 x_2 y_2-x_1 y_3-x_3 y_1
	\end{pmatrix} , \\
(X \otimes Y)_{3a} &\sim & \begin{pmatrix}
       x_2 y_3-x_3 y_2 \\
        x_1 y_2-x_2 y_1 \\
        x_3 y_1-x_1 y_3
\end{pmatrix}.
\end{eqnarray}
These relations are used in the  construction of the mass matrices given in Eq. (\ref{eq:cl-mass-matrix}) and Eq. (\ref{eq:total-mass-matrix}). 

\acknowledgments
This work has been supported in part by the Polish National Science Center (NCN) under grant 2020/37/B/ST2/02371, the Freedom of Research (Swoboda Badań) and the Research Excellence Initiative of the University of Silesia in Katowice. BK would like to thank José W. F. Valle and Claudia Hagedorn for useful discussions. BK also acknowledges hospitality at the Korea Institute of Advanced Study, Seoul, where part of this work has been completed.  SM acknowledges the financial support from the National Research Foundation of Korea grant 2022R1A2C1005050.

	\bibliography{bibliography,ref1}
	\bibliographystyle{utphys}

\end{document}